\documentclass[11pt]{article}

\usepackage[top=0.7in, bottom=1in, left=1in, right=1in]{geometry}
\usepackage{amsmath, amssymb, amsthm}
\usepackage{graphicx}
\usepackage{natbib}
\usepackage{hyperref}
\usepackage{setspace}
\usepackage{authblk}
\usepackage{bm}
\usepackage[ruled]{algorithm2e}
\usepackage{algpseudocode}
\usepackage{booktabs}
\usepackage{multirow} 
\title{Improving Transportability of Regression Calibration Under the Main/External Validation Study Design}
\author[1]{Zexiang Li}
\author[1,2]{Donna Spiegelman}
\author[3,4]{Molin Wang}
\author[1,5]{Zuoheng Wang}
\author[1,2]{Xin Zhou}
\affil[1]{Department of Biostatistics, Yale School of Public Health, New Haven, CT}
\affil[2]{Center on Methods for Implementation and Prevention Science, Yale School of Public Health, New Haven, CT}
\affil[3]{Department of Epidemiology, Harvard T.H. Chan School of Public Health, Boston, MA}
\affil[4]{Department of Biostatistics, Harvard T.H. Chan School of Public Health, Boston, MA}
\affil[5]{Department of Biomedical Informatics \& Data Science, Yale School of Medicine, New Haven, CT}
\date{}

\makeatletter
\renewcommand{\algocf@captiontext}[2]{#1\algocf@typo. \AlCapFnt{}#2} 
\def\@algocf@capt@plain{top}
\renewcommand{\algocf@makecaption}[2]{%
  \addtolength{\hsize}{\algomargin}%
  \sbox\@tempboxa{\algocf@captiontext{#1}{#2}}%
  \ifdim\wd\@tempboxa >\hsize
    \hskip .5\algomargin%
    \parbox[t]{\hsize}{\algocf@captiontext{#1}{#2}}
  \else%
    \global\@minipagefalse%
    \hbox to\hsize{\box\@tempboxa}
  \fi%
  \addtolength{\hsize}{-\algomargin}%
}
\makeatother
\newcommand{\tabnote}[1]{\vspace{4pt}\noindent\small #1}
\newtheorem{theorem}{Theorem}
\def\T{{ \mathrm{\scriptscriptstyle T} }}
\begin{document}

\maketitle

\begin{abstract}
In epidemiology, obtaining accurate individual exposure measurements can be costly and challenging. Thus, these measurements are often subject to error. Regression calibration with a validation study is widely employed as a study design and analysis method to correct for measurement error in the main study due to its broad applicability and simple implementation. However, relying on an external validation study to assess the measurement error process carries the risk of introducing bias into the analysis. Specifically, if the parameters of regression calibration model estimated from the external validation study are not transportable to the main study, the subsequent estimated parameter describing the exposure-disease association will be biased. In this work, we improve the regression calibration method for linear regression models using an external validation study. Unlike the original approach, our proposed method ensures that the regression calibration model is transportable by estimating the parameters in the measurement error generating process using the external validation study and obtaining the remaining parameter values in the regression calibration model directly from the main study. This guarantees that parameter values in the regression calibration model will be applicable to the main study. We derived the theoretical properties of our proposed method. The simulation results show that the proposed method effectively reduces bias and maintains nominal confidence interval coverage. We applied this method to data from the Health Professionals Follow-Up Study (main study) and the Men's Lifestyle Validation Study (external validation study) to assess the effects of dietary intake on body weight.

\noindent\textbf{Keywords:} External Validation Study; Measurement Error; Regression Calibration; Transportability
\end{abstract}


\section{Introduction}
Exposures in epidemiologic studies, such as dietary intake and physical activity, are often subject to substantial measurement error \citep{Thiebaut2007,Ferrari2007}. When the exposures measured with error, also called surrogate exposures, are used instead of the true values of the exposures, called true exposures, in an analysis, the estimates are often biased \citep{buzasMeasurementError2014}. For example, the self-reported food-frequency questionnaire is often used to measure the daily intake of foods and nutrients over the past several years \citep{willettNutritionalEpidemiology2012, yuanRelativeValidityNutrient2018}, but the measurement error in food-frequency questionnaires may lead to bias in the estimated effects \citep{freedmanDealingDietaryMeasurement2011}. Therefore, it is critical to minimize, or ideally eliminate,  the impact of measurement error on the estimated effects of true exposures.

To achieve unbiased or nearly unbiased estimates for exposures-outcome association, many approaches have been developed in the literature. These include simulation extrapolation \citep{cookSimulationExtrapolationEstimationParametric1994, Raymond1996}, regression calibration  \citep{carrollMeasurementErrorNonlinear2006, rosnerCorrectionLogisticRegression1989}, Bayesian methods \citep{Petros1995}, the conditional-score method \citep{Stefanski1987}, and the corrected-score method \citep{Nakamura1990, Stefanski1989}. Among these methods, regression calibration is widely used due to its broad applicability and simple implementation \citep{rosnerCorrectionLogisticRegression1990, rosnerCorrectionLogisticRegression1992, Pierce2004,Shaw2018}. 

The regression calibration method usually involves a main study, where surrogate exposure is measured, and a validation study, where both surrogate and true exposures are measured on the same observations. Regression calibration models the relationship between the true and surrogate exposures in the validation study, then applies the estimated model to correct for measurement error in the main study. There are two types of validation studies: internal and external. The ideal setting is with an internal validation study, conducted within the main study. However, it is often costly to conduct a new validation study to collect true exposures. When an internal validation study is not feasible, an alternative option is to use an existing external validation study,
i.e., validation data from outside the main study. 
This work is motivated by two such studies, the Women's Lifestyle Validation Study \citep{yuanRelativeValidityNutrient2018} and the Men's Lifestyle Validation Study \citep{chomistekPhysicalActivityAssessment2017}. However, the validity of estimation and inference in the main/external validation study design can be impacted by the issue of transportability \citep{carrollMeasurementErrorNonlinear2006}. 

A model is considered transportable from the external validation study to the main study if it is correct and holds with the same parameter values in both studies. The regression calibration method requires that the regression calibration model, specifically the 
regression model of the true exposures on the surrogates and confounders, is transportable. However, in practice, ``much, much more rarely, the same regression calibration model can be assumed to hold across different studies'' \citep{carrollMeasurementErrorNonlinear2006}. By contrast, in nutritional epidemiology, it is often reasonable to assume that the model for the conditional expectation of surrogate exposures, given true exposures and confounders, is transportable \citep{Keogh2020}. This assumption is referred to as the single transportability assumption \citep{Wong2020}. 

The regression calibration model is usually estimated in the external validation study. However, the single transportability assumption alone does not ensure that this calibration model is transportable to the main study. In this work, we improve the original regression calibration method in linear regression models, requiring only the single transportability assumption for full validity. Unlike the original regression calibration method, which constructs the regression calibration model solely from the external validation study data, our proposed method leverages data from both the main study and the external validation study to estimate the regression calibration model 
for the main study. We then derive the asymptotic properties of our proposed method for estimation and inference. 

The contributions of this article are as follows. First, applications of regression calibration usually assume that the regression calibration model learned from the external validation study is transportable to the main study \citep{MACMAHON1990765, rosnerCorrectionLogisticRegression1990, carrollMeasurementErrorNonlinear2006}. However, this assumption does not always hold in practice. When the regression calibration model is not transportable, there has been no solution to address this issue. This work aims to fill this gap. Second, the method developed here is semiparametric, and does not rely on specific distributional assumptions for error terms. We apply linear operators to relax any such constraints, enhancing flexibility of the proposed method. Third, for linear regression models, we show that the estimates of the proposed method are consistent. We also derive estimators for their asymptotic variance.

\section{Methods}

\subsection{Notation}

Consider a main/external validation study design with sample sizes $n_M$ and $n_V$ for the main study and the external validation study, respectively. Let $\bm{Y}$, $\bm{X}$, $\bm{Z}$, and $\bm{W}$ denote the outcome, $p$-dimensional true exposures, $p$-dimensional surrogate exposures, and $q$-dimensional confounders measured without error, respectively. We observe $(Y_i,\bm{Z}_i,\bm{W}_i)$ for each individual $i$, $i=1,\ldots, n_M$, in the main study, and $(\bm{X}_i,\bm{Z}_i,\bm{W}_i)$ for each individual $i$, $i=n_M+1,\ldots, n_M+n_V$, in the external validation study. 

\subsection{Regression Calibration}

In the absence of measurement error, we assume that, given $\bm{X}_i$ and $\bm{W}_i$,  $Y_i$ follows a linear model,
\begin{equation}
	E(Y_i\mid\bm{X}_i,\bm{W}_i)=\beta_0 + \bm{\beta}_1^\T \bm{X}_i + \bm{\beta}_2^\T \bm{W}_i, \label{eq:0}
\end{equation}
where the superscript $^\T$ represents transpose, $\beta_0$ is the intercept, $\bm{\beta}_1$ and $\bm{\beta}_2$ are $p$-dimensional and $q$-dimensional vectors of regression parameters, respectively. $\bm{\beta}_1$ is the parameter of interest. However, in our setting, since $\bm{X}_i$ is not observed in the main study, we cannot estimate $\bm{\beta}_1$ directly. Instead, surrogate exposures $\bm{Z}_i$ are observed in the main study, allowing us to fit the model,
\begin{equation}
	E({Y}_i\mid\bm{Z}_i,\bm{W}_i)=\beta_0^* + {\bm{\beta}_1^{*}}^{\T} \bm{Z}_i + {\bm{\beta}_2^{*}}^{\T} \bm{W}_i,\label{eq:1}
\end{equation}
where $\beta_0^*$ is the intercept, $\bm{\beta}_1^*$ and $\bm{\beta}_2^*$ are $p$-dimensional and $q$-dimensional vectors of regression parameters, respectively. It is well known that the estimates $\widehat{\bm{\beta}}_1^*$ in model \eqref{eq:1} are biased for $\bm{\beta}_1$ due to measurement error \citep{carrollMeasurementErrorNonlinear2006}. 

The regression calibration method uses a calibration equation, $\widehat{\bm{X}} = E(\bm{X}\mid\bm{Z}, \bm{W})$, connecting each individual's true and surrogate exposures given their confounders $\bm{W}$. 
The calibration equation is estimated by the following regression calibration model in the external validation study,
\begin{equation}
	E(\bm{X}_i\mid\bm{Z}_i, \bm{W}_i) =  \bm{\gamma}_{0} + \bm{\Gamma}_{1}^\T\bm{Z}_i + \bm{\Gamma}_{2}^\T\bm{W}_i, \label{eq:2}
\end{equation}
where $\bm{\gamma}_0$ is a $p$-dimensional vector, $\bm{\Gamma}_1$ and $\bm{\Gamma}_2$ are $p\times p$ and $q\times p$ matrices, respectively. In the literature, two related methods \citep{carrollMeasurementErrorNonlinear2006,rosnerCorrectionLogisticRegression1990}, both referred to as regression calibration, use this relationship in \eqref{eq:2} to correct for measurement error. 

\cite{carrollMeasurementErrorNonlinear2006} applied the estimates $\widehat{\bm{\gamma}}_{0}, \widehat{\bm{\Gamma}}_1$ and $\widehat{\bm{\Gamma}}_2$ from model \eqref{eq:2} directly to predict the true exposures $\widehat{\bm{X}}_i = \widehat{\bm{\gamma}}_{0} + \widehat{\bm{\Gamma}}_{1}^\T\bm{Z}_i + \widehat{\bm{\Gamma}}_{2}^\T\bm{W}_i$, for $i=1,\ldots,n_M$, in the main study. Then, the regression parameters ${\bm{\beta}}_1$ can be estimated from the following model:
\begin{equation}
	E(Y_i\mid\widehat{\bm{X}}_i,\bm{W}_i)={\beta}_0 +  {\bm{\beta}}_1^\T \widehat{\bm{X}}_i+ {\bm{\beta}}_2^\T \bm{W}_i.
    \label{eq:CarrollRC}
\end{equation}
The rationale relies on the surrogacy assumption \citep{carrollMeasurementErrorNonlinear2006}, which states that measurement error contains no extra information about the outcome beyond what is already provided by true exposures. In other words, the outcome, $Y_i$, is conditionally independent of surrogates $\bm{Z}_i$ given true exposures $\bm{X}_i$ and the confounders $\bm{W}_i$. Then, we have $E(Y_i\mid \bm{Z}_i,\bm{W}_i)=E\{E(Y_i\mid \bm{X}_i,\bm{W}_i)\mid \bm{Z}_i,\bm{W}_i\}={\beta}_0 +  {\bm{\beta}}_1^\T E({\bm{X}}_i\mid \bm{Z}_i,\bm{W}_i)+ {\bm{\beta}}_2^\T \bm{W}_i$.

Instead of predicting ${\bm{X}}_i$ by $\widehat{\bm{X}}_i$, \cite{rosnerCorrectionLogisticRegression1989} used the parameters $\bm{\Gamma}_1$ in model \eqref{eq:2} to correct $\bm{\beta}_1^*$ in model \eqref{eq:1}, thereby mitigating the effects of measurement errors. For linear regression models, \citet{spiegelmanRegressionCalibrationMethod1997} showed that,
\begin{equation}
	\beta_0^* = \beta_0 + \bm{\beta}_1^\T \bm{\gamma}_{0},   \quad \bm{\beta}_1^* = \bm{\Gamma}_1\bm{\beta}_1, \quad \textrm{and} \quad \bm{\beta}_2^* = \bm{\beta}_2 + \bm{\Gamma}_2\bm{\beta}_1. \label{eq:3}
\end{equation}
Accordingly, the estimates can be corrected for measurement error:
\begin{equation}
	\widehat{{\beta}}_0 = \widehat{{\beta}}_0^* -\widehat{\bm{\gamma}}_{0}^\T (\widehat{\bm{\Gamma}}_{1})^{-1} \widehat{\bm{\beta}}_1^* , \quad \widehat{\bm{\beta}}_1= (\widehat{\bm{\Gamma}}_{1})^{-1} \widehat{\bm{\beta}}_1^*, \quad \textrm{and} \quad  \widehat{\bm{\beta}}_2 = \widehat{\bm{\beta}}_2^* - \widehat{\bm{\Gamma}}_{2}(\widehat{\bm{\Gamma}}_{1})^{-1} \widehat{\bm{\beta}}_1^*,  \label{eq:rcestimates}
\end{equation}
where $\widehat{\bm{\gamma}}_{0}$, $\widehat{\bm{\Gamma}}_{1}$ and $\widehat{\bm{\Gamma}}_{2}$ are obtained from the regression calibration model \eqref{eq:2} in the external validation study, and $\widehat{{\beta}}_0^*$, $\widehat{\bm{\beta}}_1^*$, and $\widehat{\bm{\beta}}_2^*$ are obtained from model \eqref{eq:1} in the main study. 

In both regression calibration methods, correcting $\bm{\beta}_1$ in the main study relies on the estimate of $\bm{\Gamma}_1$ from the validation study. To obtain unbiased estimates for regression calibration, thus the estimate of $\bm{\Gamma}_1$ in model \eqref{eq:2} must be accurate. Regression calibration thus assumes that the regression calibration model \eqref{eq:2} is transportable from the validation study to the main study. The transportability assumption generally holds for internal validation studies; for external validation studies, this assumption cannot be directly tested and is often assumed to be valid in practice \citep{MACMAHON1990765, rosnerCorrectionLogisticRegression1990}. However, the regression calibration model is rarely transportable across different studies \citep{carrollMeasurementErrorNonlinear2006}. When it is not transportable, regression calibration yields biased estimates, and currently, no solution exists to address this issue.

\subsection{Transportability Assumption}

To leverage information from the external validation study, we must first identify aspects of measurement error that remain the same between the main and external validation studies. Since the measurement error generating process is likely similar across both studies, it is crucial to specify this process.

Much of the measurement error literature is based on the classical measurement error \citep{Cochran1968}, where $\bm{Z}_i=\bm{X}_i+\bm{\epsilon}_e$. Another widely used model is the Berkson measurement error, which assumes $\bm{X}_i=\bm{Z}_i+\bm{\epsilon}_x$ \citep{Berkson1950}.
In this work, we consider two general classes for the underlying measurement error process: classical-like and Berkson-like \citep{carrollMeasurementErrorNonlinear2006}. In the classical-like measurement error model, the conditional distribution of $\bm{Z}$ given $(\bm{X}, \bm{W})$, $f_{\bm{Z}\mid\bm{X},\bm{W}}(\bm{z}\mid\bm{x},\bm{w})$, is specified, with the classical measurement error model as a special case. In contrast, the Berkson-like measurement error model specifies $f_{\bm{X}\mid\bm{Z},\bm{W}}(\bm{x}\mid\bm{z},\bm{w})$, the conditional distribution of $\bm{X}$ given $(\bm{Z}, \bm{W})$, with the Berkson error being a special case. Notably, the Berkson-like measurement error model is the regression calibration model.

In practice, a general rule of thumb for deciding which of these two models represents the measurement error process is that the classical-like error model may be more appropriate when surrogate exposures are uniquely measured for each individual, while the Berkson-like error model applies when all individuals within a group share the same surrogate exposures \citep{carrollMeasurementErrorNonlinear2006}.
Hence, the measurement error process is often considered to follow a classical-like measurement error model for laboratory and objective clinical measurements, for example, the self-reported food-frequency questionnaires and physical activity questionnaires \citep{carrollMeasurementErrorNonlinear2006}.

The classical-like measurement error model is assumed throughout the rest of the paper, defined by:
\begin{equation}
	\bm{Z}_i=\bm{c}_0 + \bm{C}_1^\T \bm{X}_i + \bm{C}_2^\T \bm{W}_i + \bm{\epsilon}_e, \label{eq:4}
\end{equation}
where $\bm{c}_0$ is a $p$-dimensional vector, $\bm{C}_1$ and $\bm{C}_2$ are $p\times p$ and $q\times p$ matrices, respectively. The measurement error term $\bm{\epsilon}_e$ has mean 0 and variance $\bm{\Sigma}_e$, and is independent of $\bm{X}_i$ and $\bm{W}_i$. This model describes a situation where surrogate exposures include both random error $\bm{\epsilon}_e$ and systematic error, allowing the latter to depend on true exposures $\bm{X}_i$ and error-free confounders $\bm{W}_i$. 

When the measurement error generating process is classical-like, it may be reasonable to assume model \eqref{eq:4} is transportable between the main study and the validation study, called the single transportability assumption \citep{Wong2020}. However, regression calibration requires the Berkson-like model \eqref{eq:2} to be transportable between the two studies. We note that, within a single study, a classical-like error model can be converted into a Berkson-like error model by Bayes theorem, 
\begin{equation} \label{eq:bayes}
f_{\bm{X}\mid\bm{Z},\bm{W}}(\bm{x}\mid\bm{z},\bm{w}) = \frac{f_{\bm{Z}\mid\bm{X},\bm{W}}(\bm{z}\mid\bm{x},\bm{w})f_{\bm{X}\mid\bm{W}}(\bm{x}\mid\bm{w})}{\int_{\bm{x'}}f_{\bm{Z}\mid\bm{X},\bm{W}}(\bm{z}\mid\bm{x'},\bm{w})f_{\bm{X}\mid\bm{W}}(\bm{x'}\mid\bm{w})\,d\bm{x'}},
\end{equation}
where $f_{\bm{X}\mid\bm{W}}(\bm{x}\mid\bm{w})$ is the conditional distribution of $\bm{X}$ given $\bm{W}$. 

In a main/internal validation study design, when the validation study is a simple random sample of the main study, the joint distributions of $(\bm{X}, \bm{Z}, \bm{W})$ are the same in both studies. This enables transportability of either the classical-like or Berkson-like model between both studies. In contrast, in a main/external validation study design, the joint distributions of $(\bm{X}, \bm{Z}, \bm{W})$ may differ across studies. As shown in \eqref{eq:bayes}, if the conditional distributions of $\bm{X}$ given $\bm{W}$ in two studies are different, only model \eqref{eq:4} will be transportable between two studies. Hence, regression calibration needs the double transportability assumption, which further requires that the distributions of true exposures $\bm{X}$ given $\bm{W}$ are also the same in the main study and the validation study \citep{Wong2020}, to obtain valid estimates in the main/external validation study design. However, the double transportability assumption may not hold for external validation studies \citep{carrollMeasurementErrorNonlinear2006}. In this work, we propose transportable regression calibration for valid estimation when only the single transportability assumption holds. 

\subsection{Transportable Regression Calibration for the Main/External Validation Study Design}\label{m1}

The original regression calibration method estimates model \eqref{eq:2} within the validation study, and assumes that the parameter estimates apply to the main study. In other words, regression calibration requires that the parameter estimates of \eqref{eq:2} are valid in the main study. Let us rewrite the regression calibration model \eqref{eq:2} for the main study as 
\begin{equation}
	\bm{X}_i = \bm{\gamma}_{0} + \bm{\Gamma}_{1}^\T \bm{Z}_i + \bm{\Gamma}_{2}^\T \bm{W}_i + \bm{\epsilon}_{x}, \label{eq:7}
\end{equation}
where $\bm{\epsilon}_{x}$ has mean $\bm{0}$ and variance $\bm{\Sigma}_{x}$. Here we slightly abuse the notations of $\bm{\gamma}_{0}$, $\bm{\Gamma}_{1}$ and $\bm{\Gamma}_{2}$ to represent the parameters in the main study. Under only the single transportability assumption, the parameters in \eqref{eq:2} for the external validation study and those in \eqref{eq:7} for the main study are not identical. 

In this work, we propose an improved regression calibration method that leverages the classical-like measurement error model \eqref{eq:4}, which is transportable between the main and validation studies under the single transportability assumption. 
From the validation study, we estimate the transportable parameters in
the conditional distribution of $\bm{Z}$ given $(\bm{X},\bm{W})$ in \eqref{eq:4}. Next, we estimate the distribution of $\bm{Z}$ given $\bm{W}$ in the main study, which carries information about $\bm{X}$ in the main study. Finally, we combine both distributions to derive the regression calibration model specified in \eqref{eq:7} for the main study. 

When both \eqref{eq:4} and \eqref{eq:7} hold in the main study, we could approximate the distribution of surrogate exposures $\bm{Z}_i$ in the main study by $\bm{W}_i$, which is defined as:
\begin{equation}
	\bm{Z}_i = \bm{b}_{0} + \bm{B}_{2}^\T \bm{W}_i + \bm{\epsilon}_{z}, \label{eq:6}
\end{equation}
where $\bm{b}_{0}$ is a $p$-dimensional vector, $\bm{B}_{2}$ is a $q\times p$ matrix, and $\bm{\epsilon}_{z}$ has mean 0 and variance $\bm{\Sigma}_{z}$ and is independent of $\bm{W}_i$. Intuitively, $\bm{\epsilon}_{z}$ captures information about both measurement error and true exposures that cannot be explained by confounders. Note that, under the single transportability assumption, models \eqref{eq:7} and \eqref{eq:6} are only for the main study, while model \eqref{eq:4} applies to both studies. When all the error terms in models \eqref{eq:4}, \eqref{eq:7}, and \eqref{eq:6} follow normal distributions, it is straightforward to derive $\bm{\gamma}_{0}$, $\bm{\Gamma}_{1}$ and $\bm{\Gamma}_{2}$ in model \eqref{eq:7} using parameters in \eqref{eq:4} and \eqref{eq:6}, as shown in Supplementary Material A.1. However, the normality assumption is quite strong, especially for $\bm{\epsilon}_{z}$ in \eqref{eq:6}. For example, some nutrient intakes do not follow normal distributions \citep{subarComparativeValidationBlock2001,Subar2003}. 

In this work, we apply linear operators to relax the normality assumption. Specifically, for any pair of linear operators $\bm{L}_1$ and $\bm{L}_2$,  the following equation holds: 
\begin{equation}
	\bm{L}_1( \bm{\gamma}_{0} + \bm{\Gamma}_{1}^\T \bm{Z}_i + \bm{\Gamma}_{2}^\T \bm{W}_i + \bm{\epsilon}_{x}) + \bm{L}_2 \bm{Z}_i = \bm{L}_1\bm{X}_i + \bm{L}_2(\bm{b}_{0} + \bm{B}_{2}^\T \bm{W}_i + \bm{\epsilon}_{z}). \label{eq:8}
\end{equation}
The distribution of $\bm{Z}_i$ given $\bm{X}_i$ and $\bm{W}_i$ in the main study can then be written as
\begin{equation}
	\bm{Z}_i = (\bm{L}_1\bm{\Gamma}_{1}^\T + \bm{L}_2)^{-1} \{(\bm{L}_2\bm{b}_{0} - \bm{L}_1\bm{\gamma}_{0} ) + \bm{L}_1\bm{X}_i + (\bm{L}_2\bm{B}_{2} ^\T- \bm{L}_1\bm{\Gamma}_{2}^\T)\bm{W}_i + (\bm{L}_2 \bm{\epsilon}_{z} - \bm{L}_1\bm{\epsilon}_{x})\}. \label{eq:9}
\end{equation}
Note that the relationship between $\bm{Z}_i$ and $(\bm{X}_i, \bm{W}_i)$ in \eqref{eq:9} holds for any choice of linear operators $\bm{L}_1$ and $\bm{L}_2$. Intuitively, model \eqref{eq:4} belongs to the class of linear models specified by \eqref{eq:9}, and is expected to explain the largest variation of $\bm{Z}_i$ among all models in \eqref{eq:9}. Therefore, to derive the parameters in \eqref{eq:4}, we aim to identify the linear model in \eqref{eq:9} that minimizes the variance of the error, i.e., for any $p$-dimensional vector $\bm{\alpha}$,
\begin{equation*}
	\min_{\bm{L}_1,\bm{L}_2} \bm{\alpha}^\T(\bm{L}_1\bm{\Gamma}_{1}^\T + \bm{L}_2)^{-1} (\bm{L}_2\bm{\Sigma}_{z}\bm{L}_2^\T + \bm{L}_1\bm{\Sigma}_{x}\bm{L}_1^\T)\{(\bm{L}_1\bm{\Gamma}_{1}^\T + \bm{L}_2)^{-1}\}^\T\bm{\alpha}.
\end{equation*}
The solution to this optimization problem is given by:
$$\bm{L}_1 = l\bm{\Gamma}_{1} \bm{\Sigma}_{x}^{-1}, \quad \textrm{and} \quad \bm{L}_2 = l\bm{\Sigma}_{z}^{-1}, \qquad \textrm{for any } l \in \mathbb{R}.$$
At the specific choices of $\bm{L}_1$ and $\bm{L}_2$ given above, this linear model is the same as model \eqref{eq:4}, thus we have
\begin{equation}
	\begin{aligned}
		&\bm{\gamma}_{0} = (\bm{\Sigma}_e^{-1} \bm{C}_1^\T)^{-1} (\bm{\Sigma}_{z}^{-1}\bm{b}_{0} - \bm{\Sigma}_e^{-1} \bm{c}_0),\\
		&\bm{\Gamma}_{1}^\T = (\bm{\Sigma}_e^{-1} \bm{C}_1^\T)^{-1} (\bm{\Sigma}_e^{-1} - \bm{\Sigma}_{z}^{-1}), \\
		&\bm{\Gamma}_{2}^\T = (\bm{\Sigma}_e^{-1} \bm{C}_1^\T)^{-1} (\bm{\Sigma}_{z}^{-1}\bm{B}_{2}^\T - \bm{\Sigma}_e^{-1} \bm{C}_2^\T). \label{eq:gamma}
	\end{aligned}
\end{equation}
Details for solving the optimization problem are provided in Supplementary Material A.2. The solution is identical to the one derived previously under the normality assumption of error terms, as provided in Supplementary Material A.1.

Now we have estimates for the parameters in \eqref{eq:7}. Our proposed transportable regression calibration method generates the calibration equation using model \eqref{eq:7} for the main study. Similarly to the original regression calibration, our proposed method can utilize the calibration equation in two ways, Carroll's and Rosner's. 
For Carroll's method, the true exposures can be predicted by the updated calibration equation,
\begin{eqnarray}
    \widehat{\bm{X}}_i & = & \widehat{\bm{\gamma}}_{0} + \widehat{\bm{\Gamma}}_{1}^\T\bm{Z}_i + \widehat{\bm{\Gamma}}_{2}^\T\bm{W}_i \label{eq:predictx}\\
    &=& (\widehat{\bm{\Sigma}}_e^{-1} \widehat{\bm{C}}_1^\T)^{-1}\{(\widehat{\bm{\Sigma}}_{z}^{-1}\widehat{\bm{b}}_{0} -\widehat{\bm{\Sigma}}_e^{-1}{\widehat{\bm{c}}}_0) + (\widehat{\bm{\Sigma}}_e^{-1} - \widehat{\bm{\Sigma}}_{z}^{-1})\bm{Z}_i + (\widehat{\bm{\Sigma}}_{z}^{-1}\widehat{\bm{B}}_{2}^\T -\widehat{\bm{\Sigma}}_e^{-1}\widehat{\bm{C}}_2^\T )\bm{W}_i\},\nonumber
\end{eqnarray}
where $(\widehat{\bm{c}}_0, \widehat{\bm{C}}_1, \widehat{\bm{C}}_2, \widehat{\bm{\Sigma}}_e)$ are estimated from model \eqref{eq:4} in the external validation study, and $(\widehat{\bm{b}}_{0}, \widehat{\bm{B}}_{2}, \widehat{\bm{\Sigma}}_{z})$ are estimated from model \eqref{eq:6} in the main study, both of which are obtained by least squares estimates for multiple outputs as described in \cite{Hastie2009} (See details in Supplementary Material A.3). Then the regression parameters ${\beta}_0$, $\bm{\beta}_1$ and $\bm{\beta}_2$ can be estimated through the regression model \eqref{eq:CarrollRC}. The procedure is summarized in Algorithm \ref{alg:carroll}.
\begin{algorithm}
\caption{Transportable regression calibration for Carroll's method}\label{alg:carroll}

1. $\,$ Fit the classical-like measurement error model \eqref{eq:4} to estimate ${c}_0$, ${\bm{C}}_1$, ${\bm{C}}_2$ and $\bm{\Sigma}_e$ in the validation study.

2. $\,$ Fit linear regression of $\bm{Z}$ on $\bm{W}$ of model \eqref{eq:6} to estimate ${b}_0$, ${\bm{B}}_2$ and $\bm{\Sigma}_z$ in the main study.

3. $\,$ Compute $\widehat{\bm{\gamma}}_{0}$, $\widehat{\bm{\Gamma}}_1$ and $\widehat{\bm{\Gamma}}_2$ using formula in \eqref{eq:gamma}.

4. $\,$ Predict $\widehat{\bm{X}}$ for the main study using \eqref{eq:predictx}.

5. $\,$ Fit the outcome regression model \eqref{eq:CarrollRC} to estimate ${{\beta}}_0$, ${\bm{\beta}}_1$ and ${\bm{\beta}}_2$ in the main study.


\end{algorithm}

For Rosner's method, based on \eqref{eq:rcestimates}, ${\beta}_0$, $\bm{\beta}_1$ and $\bm{\beta}_2$ can be estimated as follows,
\begin{equation}
	\begin{aligned}
		&\widehat{{\beta}}_0 = \widehat{{\beta}}_0^* - \widehat{\bm{\gamma}}_{0}^\T(\widehat{\bm{\Gamma}}_{1})^{-1} \widehat{\bm{\beta}}_1^* = \widehat{{\beta}}_0^* - (\widehat{\bm{b}}_{0}^\T\widehat{\bm{\Sigma}}_{z}^{-1} -{\widehat{\bm{c}}}_0^\T \widehat{\bm{\Sigma}}_e^{-1} ) (\widehat{\bm{\Sigma}}_e^{-1} - \widehat{\bm{\Sigma}}_{z}^{-1})^{-1}\widehat{\bm{\beta}}_1^*,\\
		&\widehat{\bm{\beta}}_1 = (\widehat{\bm{\Gamma}}_{1})^{-1} \widehat{\bm{\beta}}_1^* = \widehat{\bm{C}}_1 \widehat{\bm{\Sigma}}_e^{-1} (\widehat{\bm{\Sigma}}_e^{-1} - \widehat{\bm{\Sigma}}_{z}^{-1})^{-1}\widehat{\bm{\beta}}_1^*, \\
		&\widehat{\bm{\beta}}_2 = \widehat{\bm{\beta}}_2^* - \widehat{\bm{\Gamma}}_{2}(\widehat{\bm{\Gamma}}_{1})^{-1} \widehat{\bm{\beta}}_1^* = \widehat{\bm{\beta}}_2^* - (\widehat{\bm{B}}_{2}\widehat{\bm{\Sigma}}_{z}^{-1} -\widehat{\bm{C}}_2\widehat{\bm{\Sigma}}_e^{-1} )(\widehat{\bm{\Sigma}}_e^{-1} - \widehat{\bm{\Sigma}}_{z}^{-1})^{-1}\widehat{\bm{\beta}}_1^*. \label{eq:RSW}
	\end{aligned}
\end{equation}
The procedure is described in Algorithm \ref{alg:rosner}.
\begin{algorithm}
\caption{Transportable regression calibration for Rosner's method}\label{alg:rosner}
1. $\,$ Fit the outcome regression model \eqref{eq:1} to estimate ${\beta}^{*}_0$, ${\bm{\beta}}^{*}_1$ and ${\bm{\beta}}^{*}_2$ in the main study.

2. $\,$ Fit the classical-like measurement error model \eqref{eq:4} to estimate ${c}_0$, ${\bm{C}}_1$, ${\bm{C}}_2$ and $\bm{\Sigma}_e$ in the validation study.

3. $\,$ Fit linear regression of $\bm{Z}$ on $\bm{W}$ of model \eqref{eq:6} to estimate ${b}_0$, ${\bm{B}}_2$ and $\bm{\Sigma}_z$ in the main study.

4. $\,$ Compute $\widehat{\bm{\gamma}}_{0}$, $\widehat{\bm{\Gamma}}_1$ and $\widehat{\bm{\Gamma}}_2$ using formula in \eqref{eq:gamma}.

5. $\,$ Compute $\widehat{{\beta}}_0$, $\widehat{\bm{\beta}}_1$ and $\widehat{\bm{\beta}}_2$ using formula in \eqref{eq:RSW}.


\end{algorithm}

For the linear regression model \eqref{eq:0}, it is easy to show that the estimators by our proposed transportable Carroll's and Rosner's methods are exactly the same. Now we focus on the Rosner's method, since it provides closed-form formulas for $\widehat{{\beta}}_0$, $\widehat{\bm{\beta}}_1$ and $\widehat{\bm{\beta}}_2$ in \eqref{eq:RSW}. 

The asymptotic properties of the estimates can be presented in the following theorem.
\begin{theorem} Under the surrogacy and single transportability assumptions, and assuming models \eqref{eq:4}, \eqref{eq:7} and \eqref{eq:6} are correctly specified, $\widehat{\bm{\beta}}$ converges in probability to the true values $\bm{\beta}$, and $\surd{n_M}(\widehat{\bm{\beta}}-\bm{\beta})$ is asymptotically mean-zero multivariate normal with variance-covariance matrix that can be estimated consistently by the estimator described in Supplementary Material A.3, when both $n_M$ and $n_V$ approach infinity and $n_M/n_V \rightarrow \lambda$, where $0<\lambda<\infty$, $\bm{\beta}=({\beta}_0, \bm{\beta}_1^\T, \bm{\beta}_2^\T)^\T$ are true values of the parameters, and $\widehat{\bm{\beta}}=(\widehat{\beta}_0, \widehat{\bm{\beta}}_1^\T, \widehat{\bm{\beta}}_2^\T)^\T$.
\end{theorem}
The detailed proof is provided in Supplementary Material A.3. Here, we present an outline of the proof. Consistency can be established using the continuous mapping theorem. For asymptotic normality, note that
$(\widehat{\beta}_0^*, \widehat{\bm{\beta}}_1^*, \widehat{\bm{\beta}}_2^*)$ and $(\widehat{\bm{b}}_{0}, \widehat{\bm{B}}_{2}, \widehat{\bm{\Sigma}}_{z})$ are estimated from the main study, and $(\widehat{\bm{c}}_0, \widehat{\bm{C}}_1, \widehat{\bm{C}}_2, \widehat{\bm{\Sigma}}_e)$ are estimated from the external validation study. Hence, they are independent.
By the properties of linear regression, we have that $(\widehat{\bm{c}}_0, \widehat{\bm{C}}_1, \widehat{\bm{C}}_2)$ are independent of $\widehat{\bm{\Sigma}}_e$, and $(\widehat{\bm{b}}_{0}, \widehat{\bm{B}}_{2})$ are independent of $\widehat{\bm{\Sigma}}_{z}$. 
Furthermore, it can be shown that $(\widehat{\beta}_0^*, \widehat{\bm{\beta}}_1^*, \widehat{\bm{\beta}}_2^*)$ and $(\widehat{\bm{b}}_{0}, \widehat{\bm{B}}_{2}, \widehat{\bm{\Sigma}}_{z})$ are asymptotically uncorrelated. The asymptotic normality follows from the multivariate delta method.

\section{Simulation Studies}

We carried out extensive simulation studies to evaluate the performance of the proposed transportable regression calibration method under the main/external validation study design, with sample sizes of $n_M=10,000$ for the main study and $n_V=500$ for the external validation study. In all simulations, $W$ included a single error-free confounder generated from a normal distribution $\mathcal{N}(1,1)$. For true exposures $\bm{X}$, we considered two cases, single-exposure and multiple-exposure with four components. The distribution of $\bm{X}$ in the main study is specified by $\bm{X}_i = \bm{a}_0 + \bm{A}_2^\T \bm{W}_i + \bm{\epsilon}$, where $\bm{\epsilon}$ has mean 0 and variance $\bm{\Sigma}$. For the error term, $\bm{\epsilon}$, we considered two options, a normal distribution and a gamma distribution. These two error distributions have a mean of 0 and the same variance, but significantly different shapes. The plots of these distributions for the single-exposure case are shown in Supplementary Material.  


All simulation settings satisfied the single transportability assumption, i.e., the parameters $\bm{c}_0$, $\bm{C}_1$, $\bm{C}_2$ and $\bm{\Sigma}_e$ in model \eqref{eq:4} were the same in both the main study and the external validation study. 
Model \eqref{eq:4} was then used to generate surrogate exposures $\bm{Z}$ in both studies. The outcome $\bm{Y}$ was generated by model \eqref{eq:0} with random errors following a standard normal distribution.

We considered three scenarios for generating true exposures $\bm{X}$. The first scenario satisfies the double transportability assumption, while the other two do not, introducing the transportability issue. Let $(\bm{\mu}_M, \bm{\Sigma}_M)$ and $(\bm{\mu}_V, \bm{\Sigma}_V)$ denote the mean and variance of the conditional distribution of $\bm{X}$ given ${W}$ in the main study and the external validation study, respectively. Specifically, in Scenario 1, $\bm{\mu}_M=\bm{\mu}_V, \bm{\Sigma}_M=\bm{\Sigma}_V$; in Scenario 2, both the mean and variance of $\bm{X}$ in the external validation study are 20\% smaller than those in the main study ($\bm{\mu}_V = 0.8\bm{\mu}_M, \bm{\Sigma}_V=0.8\bm{\Sigma}_M$); in Scenario 3, the mean and variance in the main study are 20\% smaller than those in the external validation study ($\bm{\mu}_V = 1.25\bm{\mu}_M, \bm{\Sigma}_V = 1.25\bm{\Sigma}_M$). Additionally, smaller values were set to the off-diagonal elements of $\bm{\Sigma}_M$ and $\bm{\Sigma}_V$ for the multiple-exposure case. 

In each scenario, we examined two sets of parameters to represent small and large measurement errors, respectively. For a single-exposure $X$, a measurement error is considered small if $\beta_1^2\text{var}(X\mid Z, W)$ is less than 0.5, as suggested by \cite{Kuha1994}. Specifically, with fixed $\beta_1 = 1$, we controlled the measurement error level by varying the value of $\Sigma_e$. 
For the multiple-exposure case, we used four components in $\bm{X}$ with fixed $\bm{\beta}_1 = (1.2, 1.1, 0.9, 0.8)^\T$. Specifically, we applied the same rule for each component to determine $\bm{\Sigma}_e$ for different measurement error levels, i.e., for $j=1,\ldots,4$, $\beta_{1,j}^2\text{var}(X_{ij}\mid\bm{Z}_i, W_i)$ is less than $0.5$ or not. 

The simulations were repeated 10,000 times for each scenario. We compared our proposed transportable regression calibration method with a naive method, which uses surrogate exposures in place of true exposures, and the original regression calibration.

When the error term $\bm{\epsilon}$ was normally distributed, the simulation results for a single exposure are shown in Table~\ref{tc1}, while Tables~\ref{tc2} and \ref{tc3} show the results for the multiple-exposure case with small and large measurement errors, respectively. The naive method was severely biased in all simulation scenarios examined. In Scenario 1, which considers the double transportability assumption, both the original and transportable regression calibration methods consistently showed low bias and approximately 95\% coverage probability. The variance of our transportable regression calibration method was larger than that of regression calibration, which is expected, as our proposed method involves more parameters. It is worth noting that the efficiency loss of our proposed method compared to the original regression calibration is modest when measurement error is small, indicating that our method will be a viable option in these settings. When only the single transportability assumption was satisfied, as in Scenarios 2 and 3, the original regression calibration estimates were biased due to the transportability issue, while our proposed transportable regression calibration method addressed this issue and consistently exhibited low bias. In addition, the empirical coverage probability of the asymptotic confidence interval based on our proposed method was generally close to the nominal level of 95\%, while for the original regression calibration method, the coverage probability was not satisfactory. In all situations we studied, the average of the standard error estimates for our proposed method was close to the standard deviation of the point estimates, indicating the good performance of our variance estimator in finite samples. 

When the error term $\bm{\epsilon}$ followed the gamma distribution, the results are presented in the supplementary material. Similar findings were observed as previously, indicating that our proposed transportable regression calibration method is robust to different error distributions, as expected from the theoretical derivation.

\begin{table}[htbp]
	\centering
	\caption{Simulation results for the single-exposure case with $\beta_1 = 1$. ME is the measurement error level. $\widehat{\beta}_1$ is the estimate of $\beta_1$. Bias(\%) is the relative bias, i.e. Bias(\%)=$(\widehat{\beta}_1 - \beta_1)/\beta_1\times 100$. SE is the mean of the estimated standard error of $\widehat{\beta}_1$. SD is the empirical standard deviation of the $\widehat{\beta}_1$ values. CP is the empirical coverage probability of the asymptotic 95\% confidence interval.}
	\begin{tabular}{cccccccc}
\toprule
		& ME  & Method & $\widehat{\beta}_1$ & Bias(\%)   & SE    & SD    & CP \\ \midrule
		\multirow{6}[0]{*}{\rotatebox{90}{Scenario 1}} & \multirow{3}[0]{*}{\rotatebox{90}{Small}}   & Our proposed method & 1.00  & -0.05\% & 0.047 & 0.047 & 95.01\% \\
          &       & Regression calibration & 1.00  & 0.13\% & 0.034 & 0.035 & 94.84\% \\
          &       & Naive estimator & 0.67  & -32.89\% & 0.009 & 0.010 & 0.00\% \\ \cmidrule{2-8}
		& \multirow{3}[0]{*}{\rotatebox{90}{Large}}    & Our proposed method & 1.01  & 1.03\% & 0.150 & 0.151 & 93.84\% \\
          &       & Regression calibration & 1.00  & 0.33\% & 0.067 & 0.067 & 94.99\% \\
          &       & Naive estimator & 0.34  & -66.21\% & 0.007 & 0.007 & 0.00\% \\ \midrule
		\multirow{6}[0]{*}{\rotatebox{90}{Scenario 2}} & \multirow{3}[0]{*}{\rotatebox{90}{Small}}   & Our proposed method & 1.00  & -0.12\% & 0.049 & 0.049 & 94.84\% \\
          &       & Regression calibration & 1.08  & 8.28\% & 0.041 & 0.041 & 48.48\% \\
          &       & Naive estimator & 0.67  & -32.90\% & 0.009 & 0.009 & 0.00\% \\ \cmidrule{2-8}
		& \multirow{3}[0]{*}{\rotatebox{90}{Large}}     & Our proposed method & 1.01  & 1.13\% & 0.154 & 0.156 & 94.12\% \\
          &       & Regression calibration & 1.17  & 17.20\% & 0.087 & 0.088 & 51.22\% \\
          &       & Naive estimator & 0.34  & -66.23\% & 0.007 & 0.008 & 0.00\% \\ \midrule
		\multirow{6}[0]{*}{\rotatebox{90}{Scenario 3}} & \multirow{3}[0]{*}{\rotatebox{90}{Small}}   & Our proposed method & 1.00  & -0.04\% & 0.045 & 0.045 & 94.88\% \\
          &       & Regression calibration & 0.94  & -6.49\% & 0.029 & 0.029 & 39.97\% \\
          &       & Naive estimator & 0.67  & -32.88\% & 0.009 & 0.009 & 0.00\% \\ \cmidrule{2-8}
		& \multirow{3}[0]{*}{\rotatebox{90}{Large}}     & Our proposed method & 1.01  & 0.93\% & 0.148 & 0.153 & 93.29\% \\
          &       & Regression calibration & 0.87  & -12.90\% & 0.053 & 0.053 & 32.92\% \\
          &       & Naive estimator & 0.34  & -66.22\% & 0.007 & 0.007 & 0.00\% \\
  \bottomrule
	\end{tabular}%
	\label{tc1}%
        \tabnote{\\Scenario 1 satisfies the double transportability assumption, i.e. $\bm{\mu}_M=\bm{\mu}_V, \bm{\Sigma}_M=\bm{\Sigma}_V$. While Scenarios 2 and 3 only meet the single transportability assumption, i.e. $\bm{\mu}_V = 0.8\bm{\mu}_M, \bm{\Sigma}_V=0.8\bm{\Sigma}_M$ for Scenario 2 and $\bm{\mu}_V = 1.25\bm{\mu}_M, \bm{\Sigma}_V = 1.25\bm{\Sigma}_M$ for Scenario 3. }
\end{table}%

\begin{table}[htbp]
	\centering
	\caption{Simulation results for the multiple-exposure case with small measurement errors. $\beta$ is the true value of the corresponding component of $\bm{\beta}_1$. 
    $\widehat{\beta}$ is the estimate of $\beta$. Bias(\%) is the relative bias, i.e. Bias(\%)=$(\widehat{\beta} - \beta)/\beta\times 100$. SE is the mean of the estimated standard error of $\widehat{\beta}$. SD is the empirical standard deviation of the $\widehat{\beta}$ values. CP is the empirical coverage probability of the asymptotic 95\% confidence interval.}
	\begin{tabular}{ccccccccc}
 \toprule
		&  & $\beta$  & Method & $\widehat{\beta}$ & Bias(\%)   & SE    & SD    & CP \\
  \midrule
		\multirow{12}[0]{*}{\rotatebox{90}{Scenario 1}} & $\beta_{1,1}$ & 1.2   & Our proposed method & 1.20  & -0.16\% & 0.077 & 0.077 & 94.81\% \\
          &       &       & Regression calibration & 1.20  & 0.04\% & 0.066 & 0.066 & 94.92\% \\
          &       &       & Naive estimator & 0.73  & -39.08\% & 0.013 & 0.013 & 0.00\% \\
  \cmidrule{2-9}
		& $\beta_{1,2}$ & 1.1   & Our proposed method & 1.10  & -0.23\% & 0.077 & 0.076 & 94.98\% \\
          &       &       & Regression calibration & 1.10  & -0.03\% & 0.066 & 0.066 & 94.82\% \\
          &       &       & Naive estimator & 0.66  & -39.79\% & 0.013 & 0.013 & 0.00\% \\
  \cmidrule{2-9}
		& $\beta_{1,3}$ & 0.9   & Our proposed method & 0.90  & -0.07\% & 0.076 & 0.075 & 95.44\% \\
          &       &       & Regression calibration & 0.90  & 0.14\% & 0.066 & 0.066 & 95.23\% \\
          &       &       & Naive estimator & 0.52  & -41.70\% & 0.013 & 0.013 & 0.00\% \\
  \cmidrule{2-9}
		& $\beta_{1,4}$ & 0.8   & Our proposed method & 0.80  & -0.13\% & 0.075 & 0.075 & 95.46\% \\
          &       &       & Regression calibration & 0.80  & 0.00\% & 0.066 & 0.066 & 94.88\% \\
          &       &       & Naive estimator & 0.46  & -42.97\% & 0.013 & 0.013 & 0.00\% \\
  \midrule
		\multirow{12}[0]{*}{\rotatebox{90}{Scenario 2}} & $\beta_{1,1}$ & 1.2   & Our proposed method & 1.20  & -0.14\% & 0.082 & 0.083 & 94.79\% \\
          &       &       & Regression calibration & 1.28  & 6.43\% & 0.078 & 0.080 & 83.99\% \\
          &       &       & Naive estimator & 0.73  & -39.09\% & 0.013 & 0.013 & 0.00\% \\
  \cmidrule{2-9}
		& $\beta_{1,2}$ & 1.1   & Our proposed method & 1.10  & -0.10\% & 0.082 & 0.082 & 94.94\% \\
          &       &       & Regression calibration & 1.17  & 6.19\% & 0.078 & 0.079 & 87.11\% \\
          &       &       & Naive estimator & 0.66  & -39.77\% & 0.013 & 0.013 & 0.00\% \\
  \cmidrule{2-9}
		& $\beta_{1,3}$ & 0.9   & Our proposed method & 0.90  & -0.02\% & 0.081 & 0.080 & 95.46\% \\
          &       &       & Regression calibration & 0.95  & 5.45\% & 0.078 & 0.078 & 90.94\% \\
          &       &       & Naive estimator & 0.52  & -41.68\% & 0.013 & 0.013 & 0.00\% \\
  \cmidrule{2-9}
		& $\beta_{1,4}$ & 0.8   & Our proposed method & 0.80  & -0.17\% & 0.080 & 0.081 & 95.15\% \\
          &       &       & Regression calibration & 0.84  & 4.66\% & 0.078 & 0.079 & 92.41\% \\
          &       &       & Naive estimator & 0.46  & -42.97\% & 0.013 & 0.013 & 0.00\% \\
  \midrule
		\multirow{12}[0]{*}{\rotatebox{90}{Scenario 3}} & $\beta_{1,1}$ & 1.2   & Our proposed method & 1.20  & -0.18\% & 0.073 & 0.073 & 94.87\% \\
          &       &       & Regression calibration & 1.14  & -5.04\% & 0.056 & 0.056 & 79.02\% \\
          &       &       & Naive estimator & 0.73  & -39.08\% & 0.013 & 0.013 & 0.00\% \\
  \cmidrule{2-9}
		& $\beta_{1,2}$ & 1.1   & Our proposed method & 1.10  & -0.21\% & 0.072 & 0.071 & 95.17\% \\
          &       &       & Regression calibration & 1.05  & -4.84\% & 0.056 & 0.055 & 83.10\% \\
          &       &       & Naive estimator & 0.66  & -39.81\% & 0.013 & 0.013 & 0.00\% \\
  \cmidrule{2-9}
		& $\beta_{1,3}$ & 0.9   & Our proposed method & 0.90  & -0.11\% & 0.071 & 0.071 & 95.25\% \\
          &       &       & Regression calibration & 0.86  & -4.23\% & 0.056 & 0.056 & 89.02\% \\
          &       &       & Naive estimator & 0.52  & -41.67\% & 0.013 & 0.013 & 0.00\% \\
  \cmidrule{2-9}
		& $\beta_{1,4}$ & 0.8   & Our proposed method & 0.80  & -0.16\% & 0.071 & 0.070 & 95.38\% \\
          &       &       & Regression calibration & 0.77  & -3.73\% & 0.056 & 0.056 & 91.00\% \\
          &       &       & Naive estimator & 0.46  & -42.99\% & 0.013 & 0.013 & 0.00\% \\
  \bottomrule
	\end{tabular}%
	\label{tc2}%
    \tabnote{\\Scenario 1 satisfies the double transportability assumption, i.e. $\bm{\mu}_M=\bm{\mu}_V, \bm{\Sigma}_M=\bm{\Sigma}_V$. While Scenarios 2 and 3 only meet the single transportability assumption, i.e. $\bm{\mu}_V = 0.8\bm{\mu}_M, \bm{\Sigma}_V=0.8\bm{\Sigma}_M$ for Scenario 2 and $\bm{\mu}_V = 1.25\bm{\mu}_M, \bm{\Sigma}_V = 1.25\bm{\Sigma}_M$ for Scenario 3. }
\end{table}%

\begin{table}[htbp]
  \centering
	\caption{Simulation results for the multiple-exposure case with large measurement errors. $\beta$ is the true value of the corresponding component of $\bm{\beta}_1$. $\widehat{\beta}$ is the estimate of $\beta$. Bias(\%) is the relative bias, i.e. Bias(\%)=$(\widehat{\beta} - \beta)/\beta\times 100$. SE is the mean of the estimated standard error of $\widehat{\beta}$. SD is the empirical standard deviation of the $\widehat{\beta}$ values. CP is the empirical coverage probability of the asymptotic 95\% confidence interval.}
	\begin{tabular}{ccccccccc}
		\toprule
		&  & $\beta^*$  & Method & $\widehat{\beta}$ & Bias(\%)   & SE    & SD    & CP \\
    \midrule
    \multirow{12}[0]{*}{\rotatebox{90}{Scenario 1}} & $\beta_{1,1}$ & 1.2   & Our proposed method & 1.21  & 0.66\% & 0.189 & 0.186 & 95.71\% \\
          &       &       & Regression calibration & 1.20  & 0.13\% & 0.123 & 0.124 & 94.87\% \\
          &       &       & Naive estimator & 0.49  & -59.55\% & 0.012 & 0.012 & 0.00\% \\
          \cmidrule{2-9}
          & $\beta_{1,2}$ & 1.1   & Our proposed method & 1.11  & 0.65\% & 0.186 & 0.183 & 96.19\% \\
          &       &       & Regression calibration & 1.10  & 0.17\% & 0.123 & 0.121 & 95.32\% \\
          &       &       & Naive estimator & 0.45  & -59.40\% & 0.012 & 0.012 & 0.00\% \\
          \cmidrule{2-9}
          & $\beta_{1,3}$ & 0.9   & Our proposed method & 0.90  & 0.48\% & 0.182 & 0.179 & 96.58\% \\
          &       &       & Regression calibration & 0.90  & 0.28\% & 0.123 & 0.123 & 95.26\% \\
          &       &       & Naive estimator & 0.37  & -59.07\% & 0.012 & 0.012 & 0.00\% \\
          \cmidrule{2-9}
          & $\beta_{1,4}$ & 0.8   & Our proposed method & 0.80  & 0.26\% & 0.181 & 0.179 & 96.55\% \\
          &       &       & Regression calibration & 0.80  & 0.01\% & 0.123 & 0.124 & 94.67\% \\
          &       &       & Naive estimator & 0.33  & -58.80\% & 0.012 & 0.012 & 0.00\% \\
          \midrule
    \multirow{12}[0]{*}{\rotatebox{90}{Scenario 2}} & $\beta_{1,1}$ & 1.2   & Our proposed method & 1.21  & 0.80\% & 0.196 & 0.193 & 96.01\% \\
          &       &       & Regression calibration & 1.35  & 12.84\% & 0.157 & 0.159 & 85.08\% \\
          &       &       & Naive estimator & 0.49  & -59.54\% & 0.012 & 0.012 & 0.00\% \\
          \cmidrule{2-9}
          & $\beta_{1,2}$ & 1.1   & Our proposed method & 1.11  & 0.75\% & 0.194 & 0.192 & 95.77\% \\
          &       &       & Regression calibration & 1.24  & 12.39\% & 0.157 & 0.158 & 87.57\% \\
          &       &       & Naive estimator & 0.45  & -59.40\% & 0.012 & 0.012 & 0.00\% \\
          \cmidrule{2-9}
          & $\beta_{1,3}$ & 0.9   & Our proposed method & 0.91  & 0.58\% & 0.189 & 0.184 & 96.73\% \\
          &       &       & Regression calibration & 1.01  & 11.69\% & 0.156 & 0.158 & 90.20\% \\
          &       &       & Naive estimator & 0.37  & -59.05\% & 0.012 & 0.012 & 0.00\% \\
          \cmidrule{2-9}
          & $\beta_{1,4}$ & 0.8   & Our proposed method & 0.80  & 0.11\% & 0.188 & 0.186 & 96.62\% \\
          &       &       & Regression calibration & 0.88  & 10.50\% & 0.156 & 0.158 & 91.92\% \\
          &       &       & Naive estimator & 0.33  & -58.80\% & 0.012 & 0.012 & 0.00\% \\
          \midrule
    \multirow{12}[0]{*}{\rotatebox{90}{Scenario 3}} & $\beta_{1,1}$ & 1.2   & Our proposed method & 1.21  & 0.59\% & 0.183 & 0.180 & 95.91\% \\
          &       &       & Regression calibration & 1.08  & -9.89\% & 0.098 & 0.099 & 75.28\% \\
          &       &       & Naive estimator & 0.49  & -59.55\% & 0.012 & 0.012 & 0.00\% \\
          \cmidrule{2-9}
          & $\beta_{1,2}$ & 1.1   & Our proposed method & 1.11  & 0.63\% & 0.180 & 0.176 & 96.21\% \\
          &       &       & Regression calibration & 1.00  & -9.54\% & 0.098 & 0.099 & 79.82\% \\
          &       &       & Naive estimator & 0.45  & -59.41\% & 0.012 & 0.012 & 0.00\% \\
          \cmidrule{2-9}
          & $\beta_{1,3}$ & 0.9   & Our proposed method & 0.90  & 0.45\% & 0.175 & 0.170 & 96.67\% \\
          &       &       & Regression calibration & 0.82  & -9.00\% & 0.098 & 0.100 & 86.18\% \\
          &       &       & Naive estimator & 0.37  & -59.06\% & 0.012 & 0.012 & 0.00\% \\
          \cmidrule{2-9}
          & $\beta_{1,4}$ & 0.8   & Our proposed method & 0.80  & 0.05\% & 0.174 & 0.168 & 96.94\% \\
          &       &       & Regression calibration & 0.73  & -8.36\% & 0.098 & 0.098 & 88.96\% \\
          &       &       & Naive estimator & 0.33  & -58.81\% & 0.012 & 0.012 & 0.00\% \\
          \bottomrule
    \end{tabular}%
  \label{tc3}%
    \tabnote{\\Scenario 1 satisfies the double transportability assumption, i.e. $\bm{\mu}_M=\bm{\mu}_V, \bm{\Sigma}_M=\bm{\Sigma}_V$. While Scenarios 2 and 3 only meet the single transportability assumption, i.e. $\bm{\mu}_V = 0.8\bm{\mu}_M, \bm{\Sigma}_V=0.8\bm{\Sigma}_M$ for Scenario 2 and $\bm{\mu}_V = 1.25\bm{\mu}_M, \bm{\Sigma}_V = 1.25\bm{\Sigma}_M$ for Scenario 3. }
\end{table}%

\section{Illustrative Example}

We applied our transportable regression calibration method to investigate the prospective association between body weight and dietary intake, including energy intake, alcohol intake, protein intake and total fat intake, in the Health Professionals Follow-Up Study \citep{HPFS1999}, a prospective cohort study initiated in 1986 with 51,529 male health professionals aged 40 to 75 years. The outcome of this analysis is body weight measured in 2014, and the dietary intake exposures were assessed in 2010 using a 130-item semi-quantitative food-frequency questionnaire, which is subject to substantial measurement error. Thus, we treated these dietary variables as surrogate exposures $\bm{Z}$. 

In practice, confounders are usually considered and need to be carefully selected for causal analysis with regression calibration \citep{Tang2024}. 
However, since this work focuses on methodological development, 
we only include age as a confounder $W$ in the analysis for illustration purposes, as age is nearly always a confounder in studies related to body weight, e.g. \cite{Liu2016, Liu2023}. Following the inclusion and exclusion criteria in \cite{Liu2016}, we excluded participants who reported diagnoses of cancer, cardiovascular disease (stroke and myocardial infarction), obesity (BMI $\ge 30$), or diabetes, as well as those with implausible energy intake ($<800$ or $>4200$ kcal/d) in 2010. The final analysis included 9,577 men from the Health Professionals Follow-Up Study who had complete surrogate exposures measured in 2010 and body weight in 2014. 

The Men's Lifestyle Validation Study \citep{Al-Shaar2020} was used as an external validation study to assess the measurement error process. Conducted between 2011 and 2013, it included 671 men recruited from the Health Professionals Follow-Up Study and Harvard Pilgrim Health Care. In this external validation study, total daily energy expenditure measured by doubly labeled water, and alcohol, protein, and total fat intake measured by 7-day dietary records were considered more accurate than those from food-frequency questionnaires. These measurements were treated as true exposures $\bm{X}$. The analysis included 658 men from this external validation study who had complete data on both $\bm{X}$ and $\bm{Z}$. These variables are summarized in Table \ref{basic_real}. The external validation study showed substantial measurement errors in surrogate exposures, particularly for energy and protein intake. Specifically, the correlations between true and surrogate exposures are 0.25 for energy intake, 0.8 for alcohol intake, 0.25 for protein intake, and 0.37 for total fat intake, respectively. Moreover, the surrogate exposures in the validation study exhibited notable differences in both mean and variance compared to those in the main study, suggesting a possible transportability issue for the regression calibration model estimated in the external validation study. 
 
\begin{table}[!htbp]
  \caption{Mean values (standard deviations) for weight, age, surrogate exposures, and true exposures used in data analysis. }
    \begin{tabular}{cccc}
    \toprule
    & \multicolumn{1}{c}{HPFS (n = 9,577)} & \multicolumn{2}{c}{MLVS (n = 658)} \\
    \midrule
    Weight, lbs  & 175.5 (23.0) & \multicolumn{2}{c}{-} \\
    Age, years   & 71.4 (6.8) & \multicolumn{2}{c}{66.7 (7.7)} \\
    \midrule
     & \multicolumn{1}{p{9.43em}}{Surrogate Exposures} & \multicolumn{1}{p{9.285em}}{Surrogate Exposures} & \multicolumn{1}{l}{True Exposures} \\
     \midrule
    Energy, kcal/day  & 2048.3 (628.2) & 2143.9 (712.1) & 2771.6 (437.4) \\
    Alcohol, g/day & 14.0 (16.0) & 16.6 (16.9) & 18.0 (18.5) \\
    Protein, g/day & 84.7 (28.0) & 87.4 (37.0) & 95.1 (21.4) \\
    Total Fat, g/day   & 74.9 (29.6) & 80.4 (32.9) & 87.9 (24.4) \\
    \bottomrule
    \end{tabular}%
  \label{basic_real}%
  \tabnote{\\HPFS, health professionals follow-up study; MLVS, men's lifestyle validation study.}
\end{table}%

We conducted analyses using the naive method, regression calibration method, and our proposed transportable regression calibration method. The results are presented in Table \ref{tce}. Both the naive and regression calibration methods showed a significant negative association between weight and energy intake, whereas our proposed method suggested a positive but non-significant association. For alcohol intake, our proposed method and the naive method exhibited a significant positive association with weight, while the regression calibration method yielded a non-significant association. The findings for protein intake varied across methods. The regression calibration and the naive method showed a significant positive association, whereas our proposed method indicated a significant negative association. Fat intake was positively associated with weight for all three methods, with significant results for the regression calibration and naive methods, but only marginally significant for our proposed method. The estimates from our proposed method were approximately twice as large as those from the other two methods. 

In summary, all three methods showed that individuals with higher alcohol and total fat intakes had greater mean body weight, which reinforces existing evidence \citep{Traversy2015, Hooper20}. After accounting for transportability issues between the main and validation studies, there was no independent effect of increased energy intake on weight, after adjusting for the intake of alcohol, protein and total fat. After controlling for differences in alcohol, protein and total fat intakes, only carbohydrate intake is left to vary -- thus, this result may best be interpreted as no impact of change in carbohydrate intake after controlling for changes in the other macro-nutrients \citep{Willett1986, Hu1999}. In addition, our proposed method aligns well with substantial evidence indicating that a high-protein diet is a successful strategy for preventing gaining weight \citep{Morenga_Mann_2012, LEIDY20151320S}. The substantial differences in estimates between the regression calibration and our proposed method identified a potential transportability issue between the Health Professionals Follow-Up Study and the Men's Lifestyle Validation Study. Our proposed method can address this issue for the regression calibration method.

\begin{table}[!htbp]
	\centering
	\caption{Difference in weight (pounds) for unit change in dietary intake. CI is the confidence interval. }
	\begin{tabular}{cccc}
 \toprule
		& Method & Estimate  (95\% CI)  & P-Value \\
  \midrule
		Energy & Our proposed method   & 1.5 (-3.01, 6.01)
 & 0.51 \\
		(500 kcal/day)& Regression calibration    & -6.0 (-11.19, -0.81)
 & 0.03 \\
		& Naive estimator    & -4.5 (-5.48, -3.52)
 & $<0.001$ \\
  \midrule
		Alcohol & Our proposed method   & 1.21 (0.2, 2.22)
 & 0.02 \\
		(12 g/day)& Regression calibration    & 0.73 (-0.12, 1.58)
 & 0.09 \\
		& Naive estimator    & 0.96 (0.6, 1.32)
 & $<0.001$ \\
  \midrule
		Protein & Our proposed method   & -2.35 (-4.05, -0.65)
 & 0.007 \\
		(10 g/day)& Regression calibration    & 5.81 (3.68, 7.94)
 & $<0.001$ \\
		& Naive estimator    & 1.35 (1.04, 1.66)
 & $<0.001$ \\
  \midrule
		Total Fat   & Our proposed method   & 5.88 (-0.58, 12.34)
 & 0.07 \\
		(20 g/day)& Regression calibration    & 2.58 (0.66, 4.5)
 & 0.009 \\
		& Naive estimator    & 2.48 (1.87, 3.09)
 & $<0.001$ \\
  \bottomrule
	\end{tabular}%
	\label{tce}%
\end{table}%

\section{Discussion}


Although measurement error is widely recognized as a major source of bias in statistical analysis \citep{carrollMeasurementErrorNonlinear2006}, methods for correcting this bias are infrequently applied \citep{Jurek2006, BRAKENHOFF2018}. A major barrier is the high cost of conducting a validation study. As discussed previously, an alternative option, which has frequently been used, is to use existing studies as external validation data. For example,  the recently completed Multi-Cohort Eating and Activity Study for Understanding Reporting Error (MEASURE) \citep{Kirkpatrick2022}, which includes the two motivating studies for this work, can serve as external validation studies. These data are publicly available and will facilitate the application of measurement error corrections in nutritional and physical activity epidemiology.  

The method proposed in this work provides valid estimates for the main/external validation study design, and addresses the transportability issue within the commonly used regression calibration method. Compared with the original regression calibration method, our proposed transportable regression calibration method integrates data in the main study to modify the estimated regression calibration model and requires only the single transportability assumption. Furthermore, by applying linear operators in the derivation, the method developed in this work is semiparametric and does not rely on specific distributional assumptions for error terms. Finally, we verified the consistency of our proposed method through both theoretical derivation and simulation studies. 

It should be noted that the single transportability assumption cannot be directly verified in practice. When this assumption is violated, the resulting estimates of the proposed method are likely to be biased. This can be a limitation of measurement error correction methods in the main/external validation study design, although subject matter considerations typically support the single transportability assumption. For example, in the Pooling Project of Prospective Studies of Diet and Cancer \citep{Pooling2006}, the correlations between true and surrogate nutrition intakes are very similar across various studies. 

Our proposed method can also be applied to the main/internal validation study design. When the internal validation study is a simple random sample of the main study, as is evident from Scenario 1 of the simulation studies, our proposed method is less efficient than the original regression calibration method since our method requires estimating more parameters. However, when the validation study is not a simple random sample, the original regression calibration method suffers from the transportability issue and yields biased estimates, while our proposed method addresses this issue even with the internal validation study.

Several extensions of the proposed method are worthy of further investigation. In this work, we proposed an improved regression calibration method to address transportability issues in the main/external validation study design with linear outcome models. A similar process can be applied to generalized linear models and Cox proportional hazards regression models, as for the original regression calibration method \citep{rosnerCorrectionLogisticRegression1992, spiegelmanRegressionCalibrationMethod1997}. However, it is important to note that the resulting estimates may no longer be consistent, although they have been found with the original regression calibration to be approximately so under well characterized, empirically verifiable assumptions that are usually met in practice. 

\section*{Acknowledgement}
This work was partially supported by grants from the National Institutes of Health (R03CA252808, R01CA279175, and R01ES026246).

\bibliographystyle{plainnat}

\section*{Supplementary material}

The Supplementary Material includes the derivation under normality assumption, the derivation under a more general scenario (by solving an optimization problem), proof for Theorem 1, and additional simulation studies when the error term $\bm{\epsilon}$ follows a gamma distribution. 

\section*{Appendix 1: Theoretical derivations}
\subsection*{Derivation under the normality assumption}
As in the main text, the models are
\begin{equation*}
    \begin{aligned}
        &\bm{Z}_i=\bm{c}_0 + \bm{C}_1^\T \bm{X}_i + \bm{C}_2^\T \bm{W}_i + \bm{\epsilon}_e,\\
        &\bm{X}_i = \bm{\gamma}_{0} + \bm{\Gamma}_{1}^\T \bm{Z}_i + \bm{\Gamma}_{2}^\T \bm{W}_i + \bm{\epsilon}_{x},\\
        &\bm{Z}_i = \bm{b}_{0} + \bm{B}_{2}^\T \bm{W}_i + \bm{\epsilon}_{z}.
    \end{aligned}
\end{equation*}
Here, we further assume that $\bm{\epsilon}_e \sim N_p(\bm{0}, \bm{\Sigma}_e)$, $\bm{\epsilon}_x \sim N_p(\bm{0}, \bm{\Sigma}_x)$ and $\bm{\epsilon}_z \sim N_p(\bm{0}, \bm{\Sigma}_z)$, where $N_p(\bm{\mu}, \bm{\Sigma})$ is a $p$-variate normal distribution with mean $\bm{\mu}$ and variance $\bm{\Sigma}$.

Based on Bayes theorem, we have
\begin{equation*}
f(\bm{Z}_i\mid\bm{X}_i,\bm{W}_i)\propto f(\bm{X}_i\mid\bm{Z}_i,\bm{W}_i)f(\bm{Z}_i\mid\bm{W}_i),
\end{equation*}
where $\propto$ denotes proportionality. The left-hand side is
\begin{equation*}
        \begin{aligned}
		&f(\bm{Z}_i\mid\bm{X}_i,\bm{W}_i) \propto \exp\left[-\frac{1}{2} \left\{(\bm{Z}_i - \bm{c}_0 - \bm{C}_1^\T \bm{X}_i - \bm{C}_2^\T \bm{W}_i)^\T \bm{\Sigma}_e^{-1} (\bm{Z}_i - \bm{c}_0 - \bm{C}_1^\T \bm{X}_i - \bm{C}_2^\T \bm{W}_i)\right\}\right]\\
		\propto & 
        \exp\left(-\frac{1}{2}\bm{Z}_i^\T\bm{\Sigma}_e^{-1}\bm{Z}_i\right) \exp\left(\bm{Z}_i^\T\bm{\Sigma}_e^{-1}\bm{C}_1^\T \bm{X}_i\right)\exp\left(\bm{Z}_i^\T\bm{\Sigma}_e^{-1}\bm{C}_2^\T \bm{W}_i\right)\exp\left(\bm{Z}_i^\T\bm{\Sigma}_e^{-1}\bm{c}_0\right).
	\end{aligned}
\end{equation*}
The right-hand side is
\begin{equation*}
    \begin{aligned}
    &f(\bm{X}_i\mid\bm{Z}_i,\bm{W}_i)f(\bm{Z}_i\mid\bm{W}_i)\\
   \propto & \exp\left[-\frac{1}{2} (\bm{X}_i-\bm{\gamma}_{0} - \bm{\Gamma}_{1}^\T \bm{Z}_i - \bm{\Gamma}_{2}^\T \bm{W}_i)^\T \bm{\Sigma}_x^{-1} (\bm{X}_i-\bm{\gamma}_{0} - \bm{\Gamma}_{1}^\T \bm{Z}_i - \bm{\Gamma}_{2}^\T \bm{W}_i)\right]
		\exp\left[-\frac{1}{2} (\bm{Z}_i-\bm{b}_{0} - \bm{B}_{2}^\T \bm{W}_i)^\T \bm{\Sigma}_z^{-1} (\bm{Z}_i-\bm{b}_{0} - \bm{B}_{2}^\T \bm{W}_i)\right]\\
	\propto & \exp\left\{-\frac{1}{2}\bm{Z}_i^\T(\bm{\Gamma}_{1} \bm{\Sigma}_x^{-1} \bm{\Gamma}_{1}^\T+\bm{\Sigma}_z^{-1})\bm{Z}_i\right\} \exp\left(\bm{Z}_i^\T\bm{\Gamma}_{1}\bm{\Sigma}_x^{-1}\bm{X}_i\right)
    \exp\left\{\bm{Z}_i^\T(\bm{\Sigma}_z^{-1}\bm{B}_{2}^\T-\bm{\Gamma}_{1}\bm{\Sigma}_x^{-1}\bm{\Gamma}_{2}^\T)\bm{W}_i\right\}
    \exp\left\{\bm{Z}_i^\T(\bm{\Sigma}_z^{-1}\bm{b}_0-\bm{\Gamma}_{1}\bm{\Sigma}_x^{-1}\bm{\gamma}_{0})\right\}.
	\end{aligned}
\end{equation*}
Comparing similar terms on both sides, we can derive:   
\begin{equation*}
	\begin{aligned}
		&\bm{\Sigma}_{e}^{-1}\bm{c}_{0} = \bm{\Sigma}_{z}^{-1}\bm{b}_{0} - \bm{\Gamma}_{1}\bm{\Sigma}_{x}^{-1}\bm{\gamma}_{0},\\
		&\bm{\Sigma}_{e}^{-1}\bm{C}_{1}^\T = \bm{\Gamma}_{1}\bm{\Sigma}_{x}^{-1},\\
		&\bm{\Sigma}_{e}^{-1}\bm{C}_{2}^\T = \bm{\Sigma}_{z}^{-1}\bm{B}_{2}^\T - \bm{\Gamma}_{1}\bm{\Sigma}_{x}^{-1}\bm{\Gamma}_{2}^\T,\\
		&\bm{\Sigma}_{e}^{-1} = \bm{\Gamma}_{1}\bm{\Sigma}_{x}^{-1}\bm{\Gamma}_{1}^\T + \bm{\Sigma}_{z}^{-1}, 
	\end{aligned}
\end{equation*}
which implies that
\begin{equation*}
	\begin{aligned}
		&\bm{c}_{0} = (\bm{\Gamma}_{1}\bm{\Sigma}_{x}^{-1}\bm{\Gamma}_{1}^\T + \bm{\Sigma}_{z}^{-1})^{-1}(\bm{\Sigma}_{z}^{-1}\bm{b}_{0} - \bm{\Gamma}_{1}\bm{\Sigma}_{x}^{-1}\bm{\gamma}_{0}),\\
		&\bm{C}_{1}^\T = (\bm{\Gamma}_{1}\bm{\Sigma}_{x}^{-1}\bm{\Gamma}_{1}^\T + \bm{\Sigma}_{z}^{-1})^{-1}(\bm{\Gamma}_{1}\bm{\Sigma}_{x}^{-1}),\\
		&\bm{C}_{2}^\T = (\bm{\Gamma}_{1}\bm{\Sigma}_{x}^{-1}\bm{\Gamma}_{1}^\T + \bm{\Sigma}_{z}^{-1})^{-1}(\bm{\Sigma}_{z}^{-1}\bm{B}_{2}^\T - \bm{\Gamma}_{1}\bm{\Sigma}_{x}^{-1}\bm{\Gamma}_{2}^\T),\\
		&\bm{\Sigma}_{e} = (\bm{\Gamma}_{1}\bm{\Sigma}_{x}^{-1}\bm{\Gamma}_{1}^\T + \bm{\Sigma}_{z}^{-1})^{-1}. 
	\end{aligned}
\end{equation*}
Then treating $\bm{\gamma}_{0}$, $\bm{\Gamma}_{1}$, and $\bm{\Gamma}_{2}$ as unknown parameters, we can solve the above equations to get
\begin{equation*}
	\begin{aligned}
		&\bm{\gamma}_{0} = (\bm{\Sigma}_e^{-1} \bm{C}_1^\T)^{-1} (\bm{\Sigma}_{z}^{-1}\bm{b}_{0} - \bm{\Sigma}_e^{-1} \bm{c}_0),\\
		&\bm{\Gamma}_{1}^\T = (\bm{\Sigma}_e^{-1} \bm{C}_1^\T)^{-1} (\bm{\Sigma}_e^{-1} - \bm{\Sigma}_{z}^{-1}), \\
		&\bm{\Gamma}_{2}^\T = (\bm{\Sigma}_e^{-1} \bm{C}_1^\T)^{-1} (\bm{\Sigma}_{z}^{-1}\bm{B}_{2}^\T - \bm{\Sigma}_e^{-1} \bm{C}_2^\T).
	\end{aligned}
\end{equation*}

\subsection*{Optimization problem}
The solution of
\begin{equation*}
	\min_{\bm{L}_1,\bm{L}_2} \bm{\alpha}^\T(\bm{L}_1\bm{\Gamma}_{1}^\T + \bm{L}_2)^{-1} (\bm{L}_2\bm{\Sigma}_{z}\bm{L}_2^\T + \bm{L}_1\bm{\Sigma}_{x}\bm{L}_1^\T)\{(\bm{L}_1\bm{\Gamma}_{1}^\T + \bm{L}_2)^{-1}\}^\T\bm{\alpha}
\end{equation*}
is 
$$\bm{L}_1 = l\bm{\Gamma}_{1} \bm{\Sigma}_{x}^{-1}, \quad \bm{L}_2 = l\bm{\Sigma}_{z}^{-1}, \quad l \in \mathbb{R}. $$
\begin{proof}
	Let's define
	\begin{equation*}
		f = \bm{ABA}^\T = \text{tr}(\bm{ABA}^\T), \text{ where }\bm{A} = \bm{\alpha}^\T(\bm{L}_1\bm{\Gamma}_{1}^\T + \bm{L}_2)^{-1}, \bm{B} = (\bm{L}_2\bm{\Sigma}_{z}\bm{L}_2^\T + \bm{L}_1\bm{\Sigma}_{x}\bm{L}_1^\T).
	\end{equation*}
	
	Then we have
	\begin{equation}
		\begin{aligned}
			df &= \text{tr}\{(d\bm{A})\bm{BA}^\T\} + \text{tr}\{\bm{A}(d\bm{B})\bm{A}^\T\} + \text{tr}\{\bm{AB}(d\bm{A})^\T\}\\
			&= \text{tr}\{(\bm{B}+\bm{B}^\T)\bm{A}^Td\bm{A}\} + \text{tr}\{\bm{A}^\T\bm{A}(d\bm{B})\}. \label{a1}
		\end{aligned}
	\end{equation}
	
	Based on the equation \eqref{a1} and the definition of differential which is
	\begin{equation}
		df = \text{tr}\left(\frac{\partial f^\T}{\partial \bm{A}}d\bm{A}\right) + \text{tr}\left(\frac{\partial f^\T}{\partial \bm{B}}d\bm{B}\right),\label{a2}
	\end{equation}
	we have
	\begin{equation*}
		\frac{\partial f^\T}{\partial\bm{A}} = (\bm{B}+\bm{B}^\T)\bm{A}^\T,\quad \frac{\partial f^\T}{\partial\bm{B}} = \bm{A}^\T\bm{A}.\label{a3}
	\end{equation*}
	
	Plugging $\bm{A} = \bm{\alpha}^\T(\bm{L}_1\bm{\Gamma}_{1}^\T + \bm{L}_2)^{-1}$ and $\bm{B} = (\bm{L}_2\bm{\Sigma}_{z}\bm{L}_2^\T + \bm{L}_1\bm{\Sigma}_{x}\bm{L}_1^\T)$ into equation \eqref{a2}:
	\begin{equation*}
		\begin{aligned}
			df 
			&= \text{tr}\left(\frac{\partial f^\T}{\partial \bm{A}}d\bm{A}\right) + \text{tr}\left(\frac{\partial f^\T}{\partial \bm{B}}d\bm{B}\right)\\
			&= \text{tr}\left[\frac{\partial f^\T}{\partial \bm{A}}d\{\bm{\alpha}^\T(\bm{L}_1\bm{\Gamma}_{1}^\T + \bm{L}_2)^{-1}\}\right] + \text{tr}\left\{\frac{\partial f^\T}{\partial \bm{B}}d(\bm{L}_2\bm{\Sigma}_{z}\bm{L}_2^\T + \bm{L}_1\bm{\Sigma}_{x}\bm{L}_1^\T)\right\}\\
			&= \text{tr}[\{2\bm{\Sigma}_{x}\bm{L}_1^\T\bm{A}^\T\bm{A} - \bm{\Gamma}_{1}^\T(\bm{\alpha}^\T)^{-1}\bm{A}(\bm{B} + \bm{B}^\T)\bm{A}^\T\bm{A}\}d\bm{L}_1]\\
			&+ \text{tr}[\{2\bm{\Sigma}_{z}\bm{L}_2^\T\bm{A}^\T\bm{A} - (\bm{\alpha}^\T)^{-1}\bm{A}(\bm{B} + \bm{B}^\T)\bm{A}^\T\bm{A}\}d\bm{L}_2],
		\end{aligned}
	\end{equation*}
	then we have
	\begin{equation*}
		\begin{aligned}
			\frac{\partial f^\T}{\partial \bm{L}_1} &= 2\bm{\Sigma}_{x}\bm{L}_1^\T\bm{A}^\T\bm{A} - \bm{\Gamma}_{1}^\T(\bm{\alpha}^\T)^{-1}\bm{A}(\bm{B} + \bm{B}^\T)\bm{A}^\T\bm{A},\\
			\frac{\partial f^\T}{\partial \bm{L}_2} &=
			2\bm{\Sigma}_{z}\bm{L}_2^\T\bm{A}^\T\bm{A} - (\bm{\alpha}^\T)^{-1}\bm{A}(\bm{B} + \bm{B}^\T)\bm{A}^\T\bm{A}.
		\end{aligned}
	\end{equation*}
	Based on the Supremum and Infimum principle, $f$ has a minimum value, after set 
	$$\frac{\partial f^\T}{\partial \bm{L}_1} = \frac{\partial f^\T}{\partial \bm{L}_2} = \bm{0},$$
	we get the solution
	$$\bm{L}_1 = l\bm{\Gamma}_{1} \bm{\Sigma}_{x}^{-1}, \quad \bm{L}_2 =l \bm{\Sigma}_{z}^{-1}, $$
    where $l\in \mathbb{R}$. 
\end{proof}

Then according to the corresponding relationship of each term in the following equations in the main text,
\begin{equation*}
    \begin{aligned}
        &\bm{Z}_i=\bm{c}_0 + \bm{C}_1^\T \bm{X}_i + \bm{C}_2^\T \bm{W}_i + \bm{\epsilon}_e,\\
        &\bm{Z}_i = (\bm{L}_1\bm{\Gamma}_{1}^\T + \bm{L}_2)^{-1} \{(\bm{L}_2\bm{b}_{0} - \bm{L}_1\bm{\gamma}_{0} ) + \bm{L}_1\bm{X}_i + (\bm{L}_2\bm{B}_{2} ^\T- \bm{L}_1\bm{\Gamma}_{2}^\T)\bm{W}_i + (\bm{L}_2 \bm{\epsilon}_{z} - \bm{L}_1\bm{\epsilon}_{x})\},
    \end{aligned}
\end{equation*}
we have
\begin{equation*}
	\begin{aligned}
		&\bm{c}_{0} = (\bm{L}_1\bm{\Gamma}_{1}^\T + \bm{L}_2)^{-1}(\bm{L}_2\bm{b}_{0} - \bm{L}_1\bm{\gamma}_{0} ),\\
		&\bm{C}_{1}^\T\bm{X}_i = (\bm{L}_1\bm{\Gamma}_{1}^\T + \bm{L}_2)^{-1}\bm{L}_1\bm{X}_i,\\
		&\bm{C}_2^\T \bm{W}_i = (\bm{L}_1\bm{\Gamma}_{1}^\T + \bm{L}_2)^{-1}(\bm{L}_2\bm{B}_{2} ^\T- \bm{L}_1\bm{\Gamma}_{2}^\T)\bm{W}_i,\\
		&\bm{\Sigma}_{e} = (\bm{L}_1\bm{\Gamma}_{1}^\T + \bm{L}_2)^{-1} (\bm{L}_2\bm{\Sigma}_{z}\bm{L}_2^\T + \bm{L}_1\bm{\Sigma}_{x}\bm{L}_1^\T)\{(\bm{L}_1\bm{\Gamma}_{1}^\T + \bm{L}_2)^{-1}\}^\T.
	\end{aligned}
\end{equation*}
Then treating $\bm{\gamma}_{0}, \bm{\Gamma}_{1}^\T$, and $\bm{\Gamma}_{2}^\T$ as unknown parameters and substituting $\bm{L}_1$ and $\bm{L}_2$ into the above equations, we can solve the above equations to get
\begin{equation*}
	\begin{aligned}
		&\bm{\gamma}_{0} = (\bm{\Sigma}_e^{-1} \bm{C}_1^\T)^{-1} (\bm{\Sigma}_{z}^{-1}\bm{b}_{0} - \bm{\Sigma}_e^{-1} \bm{c}_0),\\
		&\bm{\Gamma}_{1}^\T = (\bm{\Sigma}_e^{-1} \bm{C}_1^\T)^{-1} (\bm{\Sigma}_e^{-1} - \bm{\Sigma}_{z}^{-1}), \\
		&\bm{\Gamma}_{2}^\T = (\bm{\Sigma}_e^{-1} \bm{C}_1^\T)^{-1} (\bm{\Sigma}_{z}^{-1}\bm{B}_{2}^\T - \bm{\Sigma}_e^{-1} \bm{C}_2^\T).
	\end{aligned}
\end{equation*}

\subsection*{Proof of Theorem 1 and variance estimates}

\begin{proof}
Proof: As with the definitions in the main text, let $\bm{Y}$, $\bm{X}$, $\bm{Z}$, and $\bm{W}$ denote the outcome, $p$-dimensional true exposures, $p$-dimensional surrogate exposures, and $q$-dimensional confounders measured without error, respectively. We observe $(Y_i,\bm{Z}_i,\bm{W}_i)$ for each individual $i$, $i=1,\ldots, n_M$, in the main study; and we observe $(\bm{X}_i,\bm{Z}_i,\bm{W}_i)$ for each individual $i$, $i=n_M+1,\ldots, n_M+n_V$, in the external validation study. We define $\widetilde{\bm{X}}_i = (1, \bm{X}_i^\T, \bm{W}_i^\T)^\T, \widetilde{\bm{W}}_i = (1, \bm{W}_i^\T)^\T, \widetilde{\bm{Z}}_i = (1, \bm{Z}_i^\T, \bm{W}_i^\T)^\T, \widetilde{\bm{X}} = (\widetilde{\bm{X}}_{n_M+1},\ldots, \widetilde{\bm{X}}_{n_M+n_V})^\T, \widetilde{\bm{W}} = (\widetilde{\bm{W}}_{1},\ldots, \widetilde{\bm{W}}_{n_M})^\T$, $\bm{Z}_M = (\bm{Z}_1,\ldots, \bm{Z}_{n_M})^\T$, $\bm{Z}_V = (\bm{Z}_{n_M+1},\ldots, \bm{Z}_{n_M + n_V})^\T$, and $\widetilde{\bm{Z}} = (\widetilde{\bm{Z}}_1,\ldots,\widetilde{\bm{Z}}_{n_M})^\T$.

The models in the main text are,
\begin{eqnarray}
E({Y}_i\mid\bm{Z}_i,\bm{W}_i) & = &\beta_0^* + {\bm{\beta}_1^{*}}^{\T} \bm{Z}_i + {\bm{\beta}_2^{*}}^{\T} \bm{W}_i,\label{C1} \\
    \bm{Z}_i & = & \bm{c}_0 + \bm{C}_1^\T \bm{X}_i + \bm{C}_2^\T \bm{W}_i + \bm{\epsilon}_e,\label{C2} \\
    \bm{Z}_i & = & \bm{b}_{0} + \bm{B}_{2}^\T \bm{W}_i + \bm{\epsilon}_{z}.\label{C3}
\end{eqnarray}

Then based on the multiple outputs' conclusion of linear regression models in \cite{Hastie2009}, we have
\begin{eqnarray*}
        (\widehat{\bm{c}}_0, \widehat{\bm{C}}_1^\T, \widehat{\bm{C}}_2^\T)^\T & = & (\widetilde{\bm{X}}^\T\widetilde{\bm{X}})^{-1}\widetilde{\bm{X}}^\T\bm{Z}_V,\\
        \widehat{\bm{\Sigma}}_e & = & n_V^{-1}(\bm{Z}_V - \widetilde{\bm{X}}(\widetilde{\bm{X}}^\T\widetilde{\bm{X}})^{-1}\widetilde{\bm{X}}^\T\bm{Z}_V)^\T(\bm{Z}_V - \widetilde{\bm{X}}(\widetilde{\bm{X}}^\T\widetilde{\bm{X}})^{-1}\widetilde{\bm{X}}^\T\bm{Z}_V),\\
        (\widehat{\bm{b}}_0, \widehat{\bm{B}}_2^\T)^\T & = & (\widetilde{\bm{W}}^\T\widetilde{\bm{W}})^{-1}\widetilde{\bm{W}}^\T\bm{Z}_M,\\
        \widehat{\bm{\Sigma}}_z & = & n_M^{-1}(\bm{Z}_M - \widetilde{\bm{W}}(\widetilde{\bm{W}}^\T\widetilde{\bm{W}})^{-1}\widetilde{\bm{W}}^\T\bm{Z}_M)^\T(\bm{Z}_M - \widetilde{\bm{W}}(\widetilde{\bm{W}}^\T\widetilde{\bm{W}})^{-1}\widetilde{\bm{W}}^\T\bm{Z}_M),\\
\end{eqnarray*}
and similarly, for the linear regression outcome model, we have,
\begin{equation*}
        (\widehat{\beta}_0^*, {\widehat{\bm{\beta}}_1^{*\T}}, {\widehat{\bm{\beta}}_2^{*\T}})^\T = (\widetilde{\bm{Z}}^\T\widetilde{\bm{Z}})^{-1}\widetilde{\bm{Z}}^\T\bm{Y}.
\end{equation*}
Consistency and asymptotic normality hold for these parameter estimates. In other words, 
\begin{equation*}
    \surd{ n_M} \left(\begin{array}{c}
        \widehat{{\beta}}_0^* - {\beta}_0^* \\
        \widehat{\bm{\beta}}_1^* - \bm{\beta}_1^* \\
        \widehat{\bm{\beta}}_2^* - \bm{\beta}_2^* \\
        \widehat{\bm{b}}_0 - \bm{b}_0 \\
        \text{vec}(\widehat{\bm{B}}_2^\T) - \text{vec}(\bm{B}_2^\T) \\
        \text{vec}(\widehat{\bm{\Sigma}}_{z})-\text{vec}(\bm{\Sigma}_{z})
    \end{array}\right) \xrightarrow[d]{} N(\bm{0}, \bm{\Sigma}_M),
\end{equation*}
where $vec()$ represents the vectorized form of a matrix. Similarly,  
\begin{equation*}
    \surd{ n_V} \left(\begin{array}{c}
        \widehat{\bm{c}}_0 - \bm{c}_0 \\
        \text{vec}(\widehat{\bm{C}}_1^\T) - \text{vec}(\bm{{C}}_1^\T) \\
        \text{vec}(\widehat{\bm{C}}_2^\T) - \text{vec}(\bm{C}_2^\T) \\
        \text{vec}(\widehat{\bm{\Sigma}}_{e})-\text{vec}(\bm{\Sigma}_{e})
    \end{array}\right) \xrightarrow[d]{} N(\bm{0}, \bm{\Sigma}_V).
\end{equation*}

Define $\bm{\theta}$ as all the parameters mentioned above, and $\widehat{\bm{\theta}}$ as their estimates. Then we can easily see that $\widehat{\bm{\theta}}$ is consistent and asymptotically normal, $\surd{n_M}(\widehat{\bm{\theta}}-\bm{\theta})\xrightarrow[d]{} N(\bm{0}, \bm{\Sigma}_D)$, where $\bm{\Sigma}_D=\left(\begin{array}{cc}\bm{\Sigma}_M & 0 \\ 0 & \lambda\bm{\Sigma}_V\end{array}\right)$, since $n_M/n_V \rightarrow \lambda$. We have derived in the main text,
\begin{equation*}
	\begin{aligned}
		&\widehat{{\beta}}_0 = \widehat{{\beta}}_0^* - \widehat{\bm{\gamma}}_{0}^\T(\widehat{\bm{\Gamma}}_{1})^{-1} \widehat{\bm{\beta}}_1^* = \widehat{{\beta}}_0^* - (\widehat{\bm{b}}_{0}^\T\widehat{\bm{\Sigma}}_{z}^{-1} -{\widehat{\bm{c}}\,}_0^\T \widehat{\bm{\Sigma}}_e^{-1} ) (\widehat{\bm{\Sigma}}_e^{-1} - \widehat{\bm{\Sigma}}_{z}^{-1})^{-1}\widehat{\bm{\beta}}_1^*,\\
		&\widehat{\bm{\beta}}_1 = (\widehat{\bm{\Gamma}}_{1})^{-1} \widehat{\bm{\beta}}_1^* = \widehat{\bm{C}}_1 \widehat{\bm{\Sigma}}_e^{-1} (\widehat{\bm{\Sigma}}_e^{-1} - \widehat{\bm{\Sigma}}_{z}^{-1})^{-1}\widehat{\bm{\beta}}_1^*, \\
		&\widehat{\bm{\beta}}_2 = \widehat{\bm{\beta}}_2^* - \widehat{\bm{\Gamma}}_{2}(\widehat{\bm{\Gamma}}_{1})^{-1} \widehat{\bm{\beta}}_1^* = \widehat{\bm{\beta}}_2^* - (\widehat{\bm{B}}_{2}\widehat{\bm{\Sigma}}_{z}^{-1} -\widehat{\bm{C}} _2\widehat{\bm{\Sigma}}_e^{-1} )(\widehat{\bm{\Sigma}}_e^{-1} - \widehat{\bm{\Sigma}}_{z}^{-1})^{-1}\widehat{\bm{\beta}}_1^*. 
	\end{aligned}
\end{equation*}
Note that the inverse of a non-singular matrix is a continuous and differentiable function of the elements of the matrix. As long as ${\bm{\Gamma}}_{1}$, ${\bm{\Sigma}}_e$, ${\bm{\Sigma}}_z$, and ${\bm{\Sigma}}_e^{-1} - {\bm{\Sigma}}_{z}^{-1}$ are non-singular, 
$\bm{\beta}=({\beta}_0, \bm{\beta}_1^\T, \bm{\beta}_2^\T)^\T$ is a differentiable function of $\bm{\theta}$, say $\bm{\beta}=\bm{h}(\bm{\beta})$, where $\bm{h}$ is differentiable. Then, the consistency of $\widehat{\bm{\beta}}$ is established by the continuous mapping theorem; The asymptotic normality is the direct application of the multivariate delta method, achieving
\begin{equation*}
\surd{n_M}(\widehat{\bm{\beta}}-\bm{\beta})\xrightarrow[d]{} N\left(\bm{0}, \frac{\partial \bm{h}(\bm{\beta})}{\bm{\beta}}\bm{\Sigma}_V\left(\frac{\partial \bm{h}(\bm{\beta})}{\bm{\beta}}\right)^\T\right).
\end{equation*}
We complete the proof.
\end{proof}


In the following, we want to specify the formula of the asymptotic variance-covariance matrix for $\surd{n_M}(\widehat{\bm{\beta}}-\bm{\beta})$. Note that $(\widehat{\beta}_0^*, \widehat{\bm{\beta}}_1^*, \widehat{\bm{\beta}}_2^*)$ and $(\widehat{\bm{b}}_{0}, \widehat{\bm{B}}_{2}, \widehat{\bm{\Sigma}}_{z})$ are estimated from the main study, and $(\widehat{\bm{c}}_0, \widehat{\bm{C}}_1, \widehat{\bm{C}}_2, \widehat{\bm{\Sigma}}_e)$ are estimated from the external validation study. Hence they are independent. Additionally, we know that $(\widehat{\bm{c}}_0, \widehat{\bm{C}}_1, \widehat{\bm{C}}_2)$ are independent of $\widehat{\bm{\Sigma}}_e$, and $(\widehat{\bm{b}}_{0}, \widehat{\bm{B}}_{2})$ are independent of $\widehat{\bm{\Sigma}}_{z}$ by the properties of linear regression estimates. Thus, we can exclude these covariance terms in our derivation. Furthermore, the derivation of the asymptotic variance-covariance matrix for $\surd{n_M}(\widehat{\bm{\beta}}-\bm{\beta})$ can be simplified by showing that $(\widehat{\beta}_0^*, \widehat{\bm{\beta}}_1^*, \widehat{\bm{\beta}}_2^*)$ and $(\widehat{\bm{b}}_{0}, \widehat{\bm{B}}_{2}, \widehat{\bm{\Sigma}}_{z})$ are asymptotically uncorrelated. Here, we show that the asymptotic covariance of $\widehat{\bm{\beta}}_1^*$ and $\widehat{\bm{\Sigma}}_{z}$ is $\bm{0}$, and other conclusions follow similarly. 
\begin{proof}
	By asymptotically uncorrelated, we mean that
	\begin{equation*}
		\surd{ n_M} \left(\begin{array}{c}
			\widehat{\bm{\beta}}_1^* - \bm{\beta}_1^* \\
			\text{vec}(\widehat{\bm{\Sigma}}_{z})-\text{vec}(\bm{\Sigma}_{z})
		\end{array}\right)
	\end{equation*}
	is asymptotically multivariate normally distributed with mean zero and a covariance matrix which is of block diagonal form. For all generalized linear models, it is well known that the estimated primary regression slope $\bm{\beta}_1^*$ has the expansion
	\begin{equation*}
		\surd{n_M} (\widehat{\bm{\beta}}_1^* - \bm{\beta}_1^*) = n_M^{-1/2} \sum_{i=1}^{n_M} \psi_1({Y}_i,\bm{Z}_i,\bm{W}_i;\bm{\beta}^*_1) + o_p(1),
	\end{equation*}
	where $\psi_1({Y}_i,\bm{Z}_i,\bm{W}_i;\bm{\beta}^*_1)$ is defined such that
	\begin{equation}
		E\{\psi_1(\bm{Y}_i,\bm{Z}_i,\bm{W}_i;\bm{\beta}^*_1)\mid(\bm{Z}_i,\bm{W}_i)\} = \bm{0}.\label{B1}
	\end{equation}
	
	Similarly, we have an expansion:
	\begin{equation}
		\surd{n_M} \{\text{vec}(\widehat{\bm{\Sigma}}_{z})-\text{vec}(\bm{\Sigma}_{z})\} = n_M^{-1/2} \sum_{i=1}^{n_M} \psi_2(\bm{Z}_i,\bm{W}_i;\bm{\Sigma}_{z}) + o_p(1)\label{B2}
	\end{equation}
	for a function $\psi_2(\bm{Z}_i,\bm{W}_i;\bm{\Sigma}_{z})$ defined such that
	\begin{equation*}
		E\{\psi_2(\bm{Z}_i,\bm{W}_i;\bm{\Sigma}_{z})\mid\bm{W}\} = \bm{0}.
	\end{equation*}
	
	We can now show that the asymptotic covariance of $\widehat{\bm{\beta}}_1^*$ and $\widehat{\bm{\Sigma}}_{z}$ is $\bm{0}$. By the theory of estimating equations, the asymptotic covariance is $\bm{F}^{-1}\bm{G}\bm{F}^{-1}$, where 
	\begin{equation*}
		\bm{F} = \left[\begin{array}{cc}
			E\left\{\frac{\partial \psi_1(\bm{Y}_i,\bm{Z}_i,\bm{W}_i;\bm{\beta}^*_1)}{\partial \bm{\beta}^*_1}\right\} & \bm{0}\\
			\bm{0} & E\left\{\frac{\partial \psi_2(\bm{Z}_i,\bm{W}_i;\bm{\Sigma}_{z})}{\partial \text{vec}(\bm{\Sigma}_{z})}\right\}
		\end{array}\right],
	\end{equation*}
	and 
	\begin{equation*}
		\bm{G} = E\left[\left\{\begin{array}{c}
			\psi_1(\bm{Y}_i,\bm{Z}_i,\bm{W}_i;\bm{\beta}^*_1)\\
			\psi_2(\bm{Z}_i,\bm{W}_i;\bm{\Sigma}_{z})
		\end{array}\right\} \left\{\begin{array}{c}
			\psi_1(\bm{Y}_i,\bm{Z}_i,\bm{W}_i;\bm{\beta}^*_1)\\
			\psi_2(\bm{Z}_i,\bm{W}_i;\bm{\Sigma}_{z})
		\end{array}\right\}^\T\right].
	\end{equation*}
	
	Based on the facts of \eqref{B1} and \eqref{B2}, we have
	\begin{equation*}
		\begin{aligned}
			E\{\psi_1(\bm{Y}_i,\bm{Z}_i,\bm{W}_i;\bm{\beta}^*_1) \psi_2^\T(\bm{Z}_i,\bm{W}_i;\bm{\Sigma}_{z})\}
			&= E[E\{\psi_1(\bm{Y}_i,\bm{Z}_i,\bm{W}_i;\bm{\beta}^*_1) \psi_2^\T(\bm{Z}_i,\bm{W}_i;\bm{\Sigma}_{z})\mid(\bm{Z}_i,\bm{W}_i)\}]\\
			&= E[E\{\psi_1(\bm{Y}_i,\bm{Z}_i,\bm{W}_i;\bm{\beta}^*_1)\mid(\bm{Z}_i,\bm{W}_i)\} \psi_2^\T(\bm{Z}_i,\bm{W}_i;\bm{\Sigma}_{z})]\\
			&= E\{\bm{0}\times \psi_2^\T(\bm{Z}_i,\bm{W}_i;\bm{\Sigma}_{z})\} = \bm{0}.
		\end{aligned}
	\end{equation*}
	Therefore, we get the fact that $E\{\psi_1(\bm{Y}_i,\bm{Z}_i,\bm{W}_i;\bm{\beta}^*_1) \psi_2^\T(\bm{Z}_i,\bm{W}_i;\bm{\Sigma}_{z})\} = \bm{0}$, which implies that the asymptotic covariance of $\widehat{\bm{\beta}}_1^*$ and $\widehat{\bm{\Sigma}}_{z}$ is $\bm{0}$.
\end{proof}

Then, to simplify the derivation, let's define:
\begin{equation*}
	\begin{aligned}
		&{\beta}_0 = {\beta}_0^* - (\bm{b}_{0}^\T{\bm{\Sigma}_{z}}^{-1} -{\bm{c}_0}^\T {\bm{\Sigma}_e}^{-1} ) ({\bm{\Sigma}_e}^{-1} - {\bm{\Sigma}_{z}}^{-1})^{-1}{\bm{\beta}_1^*} = {{\beta}_0^*} - \bm{C}\bm{B}\bm{\beta}_1^*,\\
		&\bm{\beta}_1  = {\bm{C}_1} {\bm{\Sigma}_e}^{-1} ({\bm{\Sigma}_e}^{-1} - {\bm{\Sigma}_{z}}^{-1})^{-1}{\bm{\beta}_1^*} = \bm{AB\beta}_1^*, \\
		&{\bm{\beta}_2} = {\bm{\beta}_2^*} - ({\bm{B}_{2}}{\bm{\Sigma}_{z}}^{-1} -{\bm{C}_2} {\bm{\Sigma}_e}^{-1} )({\bm{\Sigma}_e}^{-1} - {\bm{\Sigma}_{z}}^{-1})^{-1}{\bm{\beta}_1^*} = {\bm{\beta}_2^*} - \bm{DB\beta}_1^*, 
	\end{aligned}
\end{equation*}
where 
\begin{equation*}
	\begin{aligned}
		&\bm{A} = \bm{C}_1 \bm{\Sigma}_e^{-1},\\
		&\bm{B} = (\bm{\Sigma}_e^{-1} - \bm{\Sigma}_{z}^{-1})^{-1},\\
		&\bm{C} = (\bm{b}_{0}^\T{\bm{\Sigma}_{z}}^{-1} -{\bm{c}_0}^\T {\bm{\Sigma}_e}^{-1} ),  \\
		&\bm{D} = ({\bm{B}_{2}}{\bm{\Sigma}_{z}}^{-1} -{\bm{C}_2} {\bm{\Sigma}_e}^{-1} ).
	\end{aligned}
\end{equation*}
Here we slightly abuse the notations of $B$ and $C$, which differ from $(\bm{c}_0, \bm{C}_1, \bm{C}_2)$ and $(\bm{b}_{0}, \bm{B}_{2})$. 

Then we have:
\begin{equation*}
	\begin{aligned}
		&\frac{\partial (\bm{C}\bm{B}{\bm{\beta}_1^*})}{\partial \bm{C}} = (\bm{B} \bm{\beta}_1^*) \otimes \bm{I}_p, \quad \frac{\partial (\bm{C}\bm{B}{\bm{\beta}_1^*})}{\partial \bm{B}} = \bm{\beta}_1^* \otimes \bm{C}^\T,\quad \frac{\partial (\bm{C}\bm{B}{\bm{\beta}_1^*})}{\partial \bm{\beta}_1^*} =  \bm{B}^\T\bm{C}^\T,\\
		&\frac{\partial \bm{\beta}_1}{\partial \bm{A}} = (\bm{B} \bm{\beta}_1^*) \otimes \bm{I}_p, \quad 
		\frac{\partial \bm{\beta}_1}{\partial \bm{B}} = \bm{\beta}_1^* \otimes \bm{A}^\T, \quad 
		\frac{\partial \bm{\beta}_1}{\partial \bm{\beta}_1^*} = \bm{B}^\T\bm{A}^\T,\\
		&\frac{\partial (\bm{D}\bm{B}{\bm{\beta}_1^*})}{\partial \bm{D}} = (\bm{B} \bm{\beta}_1^*) \otimes \bm{I}_p, \quad \frac{\partial (\bm{D}\bm{B}{\bm{\beta}_1^*})}{\partial \bm{B}} = \bm{\beta}_1^* \otimes \bm{D}^\T,\quad \frac{\partial (\bm{D}\bm{B}{\bm{\beta}_1^*})}{\partial \bm{\beta}_1^*} =  \bm{B}^\T\bm{D}^\T,
	\end{aligned}
\end{equation*}
where $\otimes$ represents Kronecker product. 

For $\bm{A}$, we have:
$$\frac{\partial \bm{A}}{\partial \bm{C}_1^\T} = \bm{K}_{pp} (\bm{\Sigma}_e^{-1} \otimes \bm{I}_p), \quad \frac{\partial \bm{A}}{\partial \bm{\Sigma}_e} =-\{\bm{\Sigma}_e^{-1} \otimes (\bm{\Sigma}_e^{-1}\bm{C}_1^\T)\}, $$
where $\bm{K}_{pp}$ is the $p^2\times p^2$ commutation matrix. 

For $\bm{B}$, we have:
$$\frac{\partial \bm{B}}{\partial \bm{\Sigma}_e} = \{(\bm{\Sigma}_e^{-1}\bm{B}^\T) \otimes (\bm{\Sigma}_e^{-1}\bm{B})\}\bm{K}_{pp} , \quad \frac{\partial \bm{B}}{\partial \bm{\Sigma}_{z}} = -[(\bm{\Sigma}_{z}^{-1}\bm{B}^\T) \otimes \{(\bm{\Sigma}_{z}^{-1})^\T \bm{B}\}] \bm{K}_{pp}. $$

For $\bm{C}$, we have:
\begin{equation*}
	\begin{aligned}
		&\frac{\partial \bm{C}}{\partial \bm{\Sigma}_e} = \bm{K}_{pp} \{(\bm{\Sigma}_e^{-1}\bm{c}_0) \otimes \bm{\Sigma}_e^{-1}\},\\
		&\frac{\partial \bm{C}}{\partial \bm{\Sigma}_{z}} = -\bm{K}_{pp}\{(\bm{\Sigma}_{z}^{-1}\bm{b}_{0}) \otimes \bm{\Sigma}_{z}^{-1}\} ,\\
		&\frac{\partial \bm{C}}{\partial \bm{c}_0} = -\bm{K}_{pp} (\bm{I}_p \otimes \bm{\Sigma}_e^{-1}),\\
		&\frac{\partial \bm{C}}{\partial \bm{b}_{0}} = \bm{K}_{pp}(\bm{I}_p \otimes \bm{\Sigma}_{z}^{-1}).
	\end{aligned}
\end{equation*}

For $\bm{D}$, we have:
\begin{equation*}
	\begin{aligned}
		&\frac{\partial \bm{D}}{\partial \bm{\Sigma}_e} = \bm{K}_{pp} \{(\bm{\Sigma}_e^{-1}\bm{C}_2^\T) \otimes \bm{\Sigma}_e^{-1}\},\\
		&\frac{\partial \bm{D}}{\partial \bm{\Sigma}_{z}} = -\bm{K}_{pp}\{(\bm{\Sigma}_{z}^{-1}\bm{B}_{2}^\T) \otimes \bm{\Sigma}_{z}^{-1}\} ,\\
		&\frac{\partial \bm{D}}{\partial \bm{C}_2^\T} = -\bm{K}_{pp} (\bm{I}_p \otimes \bm{\Sigma}_e^{-1}),\\
		&\frac{\partial \bm{D}}{\partial \bm{B}_{2}^\T} = \bm{K}_{pp}(\bm{I}_p \otimes \bm{\Sigma}_{z}^{-1}).
	\end{aligned}
\end{equation*}

Therefore, for $\bm{\beta}_1$, we have:
\begin{equation*}
	\begin{aligned}
		\frac{\partial \bm{\beta}_1}{\partial \bm{\beta}_1^*} &=  \bm{B}^\T\bm{A}^\T,\\
		\frac{\partial \bm{\beta}_1}{\partial \bm{C}_1^\T} &= \frac{\partial \bm{A}}{\partial \bm{C}_1^\T} \frac{\partial \bm{\beta}_1}{\partial \bm{A}} = \bm{K}_{pp} (\bm{\Sigma}_e^{-1} \otimes \bm{I}_p)\{(\bm{B}\bm{\beta}_1^*)\otimes \bm{I}_p\},\\
		\frac{\partial \bm{\beta}_1}{\partial \bm{\Sigma}_e} &= \frac{\partial \bm{A}}{\partial \bm{\Sigma}_e}\frac{\partial \bm{\beta}_1}{\partial \bm{A}} + \frac{\partial \bm{B}}{\partial \bm{\Sigma}_e} \frac{\partial \bm{\beta}_1}{\partial \bm{B}} = -\{\bm{\Sigma}_e^{-1} \otimes (\bm{\Sigma}_e^{-1}\bm{C}_1^\T)\} \{(\bm{B}\bm{\beta}_1^*) \otimes \bm{I}_p\} + \{(\bm{\Sigma}_e^{-1}\bm{B}^\T) \otimes (\bm{\Sigma}_e^{-1}\bm{B})\}\bm{K}_{pp}(\bm{\beta}_1^*\otimes\bm{A}^\T),\\
		\frac{\partial \bm{\beta}_1}{\partial \bm{\Sigma}_{z}} &= \frac{\partial \bm{B}}{\partial \bm{\Sigma}_{z}}\frac{\partial \bm{\beta}_1}{\partial \bm{B}} = -[(\bm{\Sigma}_{z}^{-1}\bm{B}^\T) \otimes \{(\bm{\Sigma}_{z}^{-1})^\T\bm{B}\}]\bm{K}_{pp}(\bm{\beta}_1^*\otimes\bm{A}^\T).
	\end{aligned}
\end{equation*}

For $\bm{C}\bm{B}{\bm{\beta}_1^*}$, we have:
\begin{equation*}
	\begin{aligned}
		\frac{\partial (\bm{C}\bm{B}{\bm{\beta}_1^*})}{\partial \bm{\beta}_1^*} &= \bm{B}^\T\bm{C}^\T,\\
		\frac{\partial (\bm{C}\bm{B}{\bm{\beta}_1^*})}{\partial \bm{c}_0} &= \frac{\partial \bm{C}}{\partial \bm{c}_0} \frac{\partial (\bm{C}\bm{B}{\bm{\beta}_1^*})}{\partial \bm{C}} = -\bm{K}_{pp} (\bm{I}_p \otimes \bm{\Sigma}_e^{-1})\{(\bm{B} \bm{\beta}_1^*) \otimes \bm{I}_p\},\\
		\frac{\partial (\bm{C}\bm{B}{\bm{\beta}_1^*})}{\partial \bm{b}_{0}} &= \frac{\partial \bm{C}}{\partial \bm{b}_{0}} \frac{\partial (\bm{C}\bm{B}{\bm{\beta}_1^*})}{\partial \bm{C}} = \bm{K}_{pp}(\bm{I}_p \otimes \bm{\Sigma}_{z}^{-1})\{(\bm{B} \bm{\beta}_1^*) \otimes \bm{I}_p\},\\
		\frac{\partial (\bm{C}\bm{B}{\bm{\beta}_1^*})}{\partial \bm{\Sigma}_e} &= \frac{\partial \bm{C}}{\partial \bm{\Sigma}_e}\frac{\partial (\bm{C}\bm{B}{\bm{\beta}_1^*})}{\partial \bm{C}} + \frac{\partial \bm{B}}{\partial \bm{\Sigma}_e} \frac{\partial (\bm{C}\bm{B}{\bm{\beta}_1^*})}{\partial \bm{B}} \\
		&= \bm{K}_{pp} \{(\bm{\Sigma}_e^{-1}\bm{c}_0) \otimes \bm{\Sigma}_e^{-1}\} \{(\bm{B}\bm{\beta}_1^*) \otimes \bm{I}_p\} + \{(\bm{\Sigma}_e^{-1}\bm{B}^\T) \otimes (\bm{\Sigma}_e^{-1}\bm{B})\}\bm{K}_{pp}(\bm{\beta}_1^* \otimes \bm{C}^\T),\\
		\frac{\partial (\bm{C}\bm{B}{\bm{\beta}_1^*})}{\partial \bm{\Sigma}_{z}} &= \frac{\partial \bm{C}}{\partial \bm{\Sigma}_{z}}\frac{\partial (\bm{C}\bm{B}{\bm{\beta}_1^*})}{\partial \bm{C}} + \frac{\partial \bm{B}}{\partial \bm{\Sigma}_{z}} \frac{\partial (\bm{C}\bm{B}{\bm{\beta}_1^*})}{\partial \bm{B}} \\
		&= -\bm{K}_{pp}\{(\bm{\Sigma}_{z}^{-1}\bm{b}_{0}) \otimes \bm{\Sigma}_{z}^{-1}\} \{(\bm{B}\bm{\beta}_1^*) \otimes \bm{I}_p\} -[(\bm{\Sigma}_{z}^{-1}\bm{B}^\T) \otimes \{(\bm{\Sigma}_{z}^{-1})^\T \bm{B}\}] \bm{K}_{pp}(\bm{\beta}_1^* \otimes \bm{C}^\T).
	\end{aligned}
\end{equation*}

For $\bm{D}\bm{B}{\bm{\beta}_1^*}$, we have:
\begin{equation*}
	\begin{aligned}
		\frac{\partial (\bm{D}\bm{B}{\bm{\beta}_1^*})}{\partial \bm{\beta}_1^*} &= \bm{B}^\T\bm{D}^\T,\\
		\frac{\partial (\bm{D}\bm{B}{\bm{\beta}_1^*})}{\partial \bm{C}_2^\T} &= \frac{\partial \bm{D}}{\partial \bm{C}_2^\T} \frac{\partial (\bm{D}\bm{B}{\bm{\beta}_1^*})}{\partial \bm{D}} = -\bm{K}_{pp} (\bm{I}_p \otimes \bm{\Sigma}_e^{-1})\{(\bm{B} \bm{\beta}_1^*) \otimes \bm{I}_p\},\\
		\frac{\partial (\bm{D}\bm{B}{\bm{\beta}_1^*})}{\partial \bm{B}_{2}^\T} &= \frac{\partial \bm{D}}{\partial \bm{B}_{2}^\T} \frac{\partial (\bm{D}\bm{B}{\bm{\beta}_1^*})}{\partial \bm{D}} = \bm{K}_{pp}(\bm{I}_p \otimes \bm{\Sigma}_{z}^{-1})\{(\bm{B} \bm{\beta}_1^*) \otimes \bm{I}_p\},\\
		\frac{\partial (\bm{D}\bm{B}{\bm{\beta}_1^*})}{\partial \bm{\Sigma}_e} &= \frac{\partial \bm{D}}{\partial \bm{\Sigma}_e}\frac{\partial (\bm{D}\bm{B}{\bm{\beta}_1^*})}{\partial \bm{D}} + \frac{\partial \bm{B}}{\partial \bm{\Sigma}_e} \frac{\partial (\bm{D}\bm{B}{\bm{\beta}_1^*})}{\partial \bm{B}} \\
		&= \bm{K}_{pp} \{(\bm{\Sigma}_e^{-1}\bm{C}_2^\T) \otimes \bm{\Sigma}_e^{-1}\} \{(\bm{B}\bm{\beta}_1^*) \otimes \bm{I}_p\} + \{(\bm{\Sigma}_e^{-1}\bm{B}^\T) \otimes (\bm{\Sigma}_e^{-1}\bm{B})\}\bm{K}_{pp}(\bm{\beta}_1^* \otimes \bm{D}^\T),\\
		\frac{\partial (\bm{D}\bm{B}{\bm{\beta}_1^*})}{\partial \bm{\Sigma}_{z}} &= \frac{\partial \bm{D}}{\partial \bm{\Sigma}_{z}}\frac{\partial (\bm{D}\bm{B}{\bm{\beta}_1^*})}{\partial \bm{D}} + \frac{\partial \bm{B}}{\partial \bm{\Sigma}_{z}} \frac{\partial (\bm{D}\bm{B}{\bm{\beta}_1^*})}{\partial \bm{B}} \\
		&= -\bm{K}_{pp}\{(\bm{\Sigma}_{z}^{-1}\bm{B}_{2}^\T) \otimes \bm{\Sigma}_{z}^{-1}\} \{(\bm{B}\bm{\beta}_1^*) \otimes \bm{I}_p\} -[(\bm{\Sigma}_{z}^{-1}\bm{B}^\T) \otimes \{(\bm{\Sigma}_{z}^{-1})^\T \bm{B}\}] \bm{K}_{pp}(\bm{\beta}_1^* \otimes \bm{D}^\T).
	\end{aligned}
\end{equation*}

Thus, for the variance of $\widehat{{\beta}}_0$:
\begin{equation*}
	\begin{aligned}
		\text{var}(\widehat{{\beta}}_0) &= \text{var}(\widehat{{\beta}}_0^*)  - 2\widehat{\bm{C}}\widehat{\bm{B}}\text{cov}(\widehat{{\beta}}_0^*, \widehat{\bm{\beta}}_1^*) + \text{var}(\widehat{\bm{C}}\widehat{\bm{B}}\widehat{\bm{\beta}}_1^*)\\
		&= \text{var}(\widehat{{\beta}}_0^*)  - 2\widehat{\bm{C}}\widehat{\bm{B}}\text{cov}(\widehat{{\beta}}_0^*, \widehat{\bm{\beta}}_1^*) + \frac{\partial (\bm{C}\bm{B}\bm{\beta}_1^*)^\T}{\partial \bm{\beta}_1^*}\Bigg|_{\bm{\theta} = \widehat{\bm{\theta}}} \text{var}(\widehat{\bm{\beta}}_1^*) \frac{\partial (\bm{C}\bm{B}\bm{\beta}_1^*)}{\partial \bm{\beta}_1^*}\Bigg|_{\bm{\theta} = \widehat{\bm{\theta}}} \\
		&+ 
		\frac{\partial (\bm{C}\bm{B}{\bm{\beta}}_1^*)^\T}{\partial \bm{c}_0}\Bigg|_{\bm{\theta} = \widehat{\bm{\theta}}} \text{var}(\widehat{\bm{c}}_0) \frac{\partial (\bm{C}\bm{B}{\bm{\beta}}_1^*)}{\partial \bm{c}_0}\Bigg|_{\bm{\theta} = \widehat{\bm{\theta}}} + 
		\frac{\partial (\bm{C}\bm{B}{\bm{\beta}}_1^*)^\T}{\partial \bm{\Sigma}_e}\Bigg|_{\bm{\theta} = \widehat{\bm{\theta}}} \text{var}(\widehat{\bm{\Sigma}}_e) \frac{\partial (\bm{C}\bm{B}{\bm{\beta}}_1^*)}{\partial \bm{\Sigma}_e}\Bigg|_{\bm{\theta} = \widehat{\bm{\theta}}}\\
		&+
		\frac{\partial (\bm{C}\bm{B}{\bm{\beta}}_1^*)^\T}{\partial \bm{\Sigma}_{z}}\Bigg|_{\bm{\theta} = \widehat{\bm{\theta}}} \text{var}(\widehat{\bm{\Sigma}}_{z}) \frac{\partial (\bm{C}\bm{B}{\bm{\beta}}_1^*)}{\partial \bm{\Sigma}_{z}}\Bigg|_{\bm{\theta} = \widehat{\bm{\theta}}}+ \frac{\partial (\bm{C}\bm{B}{\bm{\beta}}_1^*)^\T}{\partial \bm{b}_{0}}\Bigg|_{\bm{\theta} = \widehat{\bm{\theta}}} \text{var}(\widehat{\bm{b}}_{0}) \frac{\partial (\bm{C}\bm{B}{\bm{\beta}}_1^*)}{\partial \bm{b}_{0}}\Bigg|_{\bm{\theta} = \widehat{\bm{\theta}}},
	\end{aligned}
\end{equation*}
where 
\begin{equation*}
	\begin{aligned}
		\frac{\partial (\bm{C}\bm{B}{\bm{\beta}_1^*})}{\partial \bm{\beta}_1^*} &= \bm{B}^\T\bm{C}^\T,\\
		\frac{\partial (\bm{C}\bm{B}{\bm{\beta}_1^*})}{\partial \bm{c}_0} &= -\bm{K}_{pp} (\bm{I}_p \otimes \bm{\Sigma}_e^{-1})\{(\bm{B} \bm{\beta}_1^*) \otimes \bm{I}_p\},\\
		\frac{\partial (\bm{C}\bm{B}{\bm{\beta}_1^*})}{\partial \bm{b}_{0}} &= \bm{K}_{pp}(\bm{I}_p \otimes \bm{\Sigma}_{z}^{-1})\{(\bm{B} \bm{\beta}_1^*) \otimes \bm{I}_p\},\\
		\frac{\partial (\bm{C}\bm{B}{\bm{\beta}_1^*})}{\partial \bm{\Sigma}_e} 
		&= \bm{K}_{pp} \{(\bm{\Sigma}_e^{-1}\bm{c}_0) \otimes \bm{\Sigma}_e^{-1}\} \{(\bm{B}\bm{\beta}_1^*) \otimes \bm{I}_p\} + \{(\bm{\Sigma}_e^{-1}\bm{B}^\T )\otimes (\bm{\Sigma}_e^{-1}\bm{B})\}\bm{K}_{pp}(\bm{\beta}_1^* \otimes \bm{C}^\T),\\
		\frac{\partial (\bm{C}\bm{B}{\bm{\beta}_1^*})}{\partial \bm{\Sigma}_{z}} 
		&= -\bm{K}_{pp}\{(\bm{\Sigma}_{z}^{-1}\bm{b}_{0}) \otimes \bm{\Sigma}_{z}^{-1}\} \{(\bm{B}\bm{\beta}_1^*) \otimes \bm{I}_p\} -[(\bm{\Sigma}_{z}^{-1}\bm{B}^\T) \otimes \{(\bm{\Sigma}_{z}^{-1})^\T \bm{B}\}] \bm{K}_{pp}(\bm{\beta}_1^* \otimes \bm{C}^\T),
	\end{aligned}
\end{equation*}
$\widehat{\bm{B}} = (\widehat{\bm{\Sigma}}_e^{-1} - \widehat{\bm{\Sigma}}_{z}^{-1})^{-1}$, $\widehat{\bm{C}} = ({\widehat{\bm{b}}_{0}}^\T{\widehat{\bm{\Sigma}}_{z}}^{-1} -{\widehat{\bm{c}}_0}^\T {\widehat{\bm{\Sigma}}_e}^{-1} )$.

For the variance of $\widehat{\bm{\beta}}_1$:
\begin{equation*}
	\begin{aligned}
		\text{var}(\widehat{\bm{\beta}}_1) &= \frac{\partial \bm{\beta}_1^\T}{\partial \bm{\beta}_1^*}\Bigg|_{\bm{\theta} = \widehat{\bm{\theta}}} \text{var}(\widehat{\bm{\beta}}_1^*) \frac{\partial \bm{\beta}_1}{\partial \bm{\beta}_1^*}\Bigg|_{\bm{\theta} = \widehat{\bm{\theta}}} + 
		\frac{\partial \bm{\beta}_1^\T}{\partial \bm{C}_1^\T}\Bigg|_{\bm{\theta} = \widehat{\bm{\theta}}} \text{var}(\widehat{\bm{C}}_1^\T) \frac{\partial \bm{\beta}_1}{\partial \bm{C}_1^\T}\Bigg|_{\bm{\theta} = \widehat{\bm{\theta}}} \\
        &+ 
		\frac{\partial \bm{\beta}_1^\T}{\partial \bm{\Sigma}_e}\Bigg|_{\bm{\theta} = \widehat{\bm{\theta}}} \text{var}(\widehat{\bm{\Sigma}}_e) \frac{\partial \bm{\beta}_1}{\partial \bm{\Sigma}_e}\Bigg|_{\bm{\theta} = \widehat{\bm{\theta}}}+
		\frac{\partial \bm{\beta}_1^\T}{\partial \bm{\Sigma}_{z}}\Bigg|_{\bm{\theta} = \widehat{\bm{\theta}}} \text{var}(\widehat{\bm{\Sigma}}_{z}) \frac{\partial \bm{\beta}_1}{\partial \bm{\Sigma}_{z}}\Bigg|_{\bm{\theta} = \widehat{\bm{\theta}}},
	\end{aligned}
\end{equation*}
where 
\begin{equation*}
	\begin{aligned}
		\frac{\partial \bm{\beta}_1}{\partial \bm{\beta}_1^*} &= \bm{B}^\T\bm{A}^\T,\\
		\frac{\partial \bm{\beta}_1}{\partial \bm{C}_1^\T} &= \bm{K}_{pp} (\bm{\Sigma}_e^{-1} \otimes \bm{I}_p)\{(\bm{B}\bm{\beta}_1^*)\otimes \bm{I}_p\},\\
		\frac{\partial \bm{\beta}_1}{\partial \bm{\Sigma}_e} &= -\{\bm{\Sigma}_e^{-1} \otimes (\bm{\Sigma}_e^{-1}\bm{C}_1^\T)\} \{(\bm{B}\bm{\beta}_1^*) \otimes \bm{I}_p\} + \{(\bm{\Sigma}_e^{-1}\bm{B}^\T) \otimes (\bm{\Sigma}_e^{-1}\bm{B})\}\bm{K}_{pp}(\bm{\beta}_1^*\otimes\bm{A}^\T),  \\
		\frac{\partial \bm{\beta}_1}{\partial \bm{\Sigma}_{z}} &= -[(\bm{\Sigma}_{z}^{-1}\bm{B}^\T) \otimes \{(\bm{\Sigma}_{z}^{-1})^\T\bm{B}\}]\bm{K}_{pp}(\bm{\beta}_1^*\otimes\bm{A}^\T).
	\end{aligned}
\end{equation*}

For the variance of $\widehat{\bm{\beta}}_2$:
\begin{equation*}
	\begin{aligned}
		\text{var}(\widehat{\bm{\beta}}_2) &= \text{var}(\widehat{\bm{\beta}}_2^*)  - 2\widehat{\bm{D}}\widehat{\bm{B}}\text{cov}(\widehat{\bm{\beta}}_2^*, \widehat{\bm{\beta}}_1^*) + \text{var}(\widehat{\bm{D}}\widehat{\bm{B}}\widehat{\bm{\beta}}_1^*)\\
		&= \text{var}(\widehat{\bm{\beta}}_2^*)  - 2\widehat{\bm{D}}\widehat{\bm{B}}\text{cov}(\widehat{\bm{\beta}}_2^*, \widehat{\bm{\beta}}_1^*) + \frac{\partial (\bm{D}\bm{B}{\bm{\beta}}_1^*)^\T}{\partial \bm{\beta}_1^*}\Bigg|_{\bm{\theta} = \widehat{\bm{\theta}}} \text{var}(\widehat{\bm{\beta}}_1^*) \frac{\partial (\bm{D}\bm{B}{\bm{\beta}}_1^*)}{\partial \bm{\beta}_1^*}\Bigg|_{\bm{\theta} = \widehat{\bm{\theta}}} \\
		&+ 
		\frac{\partial (\bm{D}\bm{B}{\bm{\beta}}_1^*)^\T}{\partial \bm{C}_2^\T}\Bigg|_{\bm{\theta} = \widehat{\bm{\theta}}} \text{var}(\widehat{\bm{C}}_2^\T) \frac{\partial (\bm{D}\bm{B}{\bm{\beta}}_1^*)}{\partial \bm{C}_2^\T}\Bigg|_{\bm{\theta} = \widehat{\bm{\theta}}} + 
		\frac{\partial (\bm{D}\bm{B}{\bm{\beta}}_1^*)^\T}{\partial \bm{\Sigma}_e}\Bigg|_{\bm{\theta} = \widehat{\bm{\theta}}} \text{var}(\widehat{\bm{\Sigma}}_e) \frac{\partial (\bm{D}\bm{B}{\bm{\beta}}_1^*)}{\partial \bm{\Sigma}_e}\Bigg|_{\bm{\theta} = \widehat{\bm{\theta}}}\\
		&+
		\frac{\partial (\bm{D}\bm{B}{\bm{\beta}}_1^*)^\T}{\partial \bm{\Sigma}_{z}}\Bigg|_{\bm{\theta} = \widehat{\bm{\theta}}} \text{var}(\widehat{\bm{\Sigma}}_{z}) \frac{\partial (\bm{D}\bm{B}{\bm{\beta}}_1^*)}{\partial \bm{\Sigma}_{z}}\Bigg|_{\bm{\theta} = \widehat{\bm{\theta}}}+ \frac{\partial (\bm{D}\bm{B}{\bm{\beta}}_1^*)^\T}{\partial \bm{B}_{2}^\T}\Bigg|_{\bm{\theta} = \widehat{\bm{\theta}}} \text{var}(\widehat{\bm{B}}_{2}^\T) \frac{\partial (\bm{D}\bm{B}{\bm{\beta}}_1^*)}{\partial \bm{B}_{2}^\T}\Bigg|_{\bm{\theta} = \widehat{\bm{\theta}}},
	\end{aligned}
\end{equation*}
where 
\begin{equation*}
	\begin{aligned}
		\frac{\partial (\bm{D}\bm{B}{\bm{\beta}_1^*})}{\partial \bm{\beta}_1^*} &= \bm{B}^\T\bm{D}^\T,\\
		\frac{\partial (\bm{D}\bm{B}{\bm{\beta}_1^*})}{\partial \bm{C}_2^\T} &= -\bm{K}_{pp} (\bm{I}_p \otimes \bm{\Sigma}_e^{-1})\{(\bm{B} \bm{\beta}_1^*) \otimes \bm{I}_p\},\\
		\frac{\partial (\bm{D}\bm{B}{\bm{\beta}_1^*})}{\partial \bm{B}_{2}^\T} &= \bm{K}_{pp}(\bm{I}_p \otimes \bm{\Sigma}_{z}^{-1})\{(\bm{B} \bm{\beta}_1^*) \otimes \bm{I}_p\},\\
		\frac{\partial (\bm{D}\bm{B}{\bm{\beta}_1^*})}{\partial \bm{\Sigma}_e} 
		&= \bm{K}_{pp} \{(\bm{\Sigma}_e^{-1}\bm{C}_2^\T) \otimes \bm{\Sigma}_e^{-1}\} \{(\bm{B}\bm{\beta}_1^*) \otimes \bm{I}_p\} + \{(\bm{\Sigma}_e^{-1}\bm{B}^\T) \otimes (\bm{\Sigma}_e^{-1}\bm{B})\}\bm{K}_{pp}(\bm{\beta}_1^* \otimes \bm{D}^\T),\\
		\frac{\partial (\bm{D}\bm{B}{\bm{\beta}_1^*})}{\partial \bm{\Sigma}_{z}} 
		&= -\bm{K}_{pp}\{(\bm{\Sigma}_{z}^{-1}\bm{B}_{2}^\T) \otimes \bm{\Sigma}_{z}^{-1}\} \{(\bm{B}\bm{\beta}_1^*) \otimes \bm{I}_p\} -[(\bm{\Sigma}_{z}^{-1}\bm{B}^\T) \otimes \{(\bm{\Sigma}_{z}^{-1})^\T \bm{B}\}] \bm{K}_{pp}(\bm{\beta}_1^* \otimes \bm{D}^\T),
	\end{aligned}
\end{equation*}
$\widehat{\bm{B}} = (\widehat{\bm{\Sigma}}_e^{-1} - \widehat{\bm{\Sigma}}_{z}^{-1})^{-1}$, and $\widehat{\bm{D}} = \left(\widehat{\bm{B}}_{2}{\widehat{\bm{\Sigma}}_{z}}^{-1} -{\widehat{\bm{C}}_2} {\widehat{\bm{\Sigma}}_e}^{-1} \right)$. 

For the covariance between $\widehat{\beta}_0$ and $\widehat{\bm{\beta}}_1$:
\begin{equation*}
	\begin{aligned}
		\text{cov}(\widehat{\beta}_0, \widehat{\bm{\beta}}_1) &= \widehat{\bm{A}}\widehat{\bm{B}}\text{cov}(\widehat{\beta}_0^*, \widehat{\bm{\beta}}_1^*)  -  \frac{\partial (\bm{C}\bm{B}{\bm{\beta}_1^*})^\T}{\partial \bm{\beta}_1^*}\Bigg|_{\bm{\theta} = \widehat{\bm{\theta}}} \text{var}(\widehat{\bm{\beta}}_1^*) \frac{\partial \bm{\beta}_1}{\partial \bm{\beta}_1^*}\Bigg|_{\bm{\theta} = \widehat{\bm{\theta}}} - 
		\frac{\partial (\bm{C}\bm{B}{\bm{\beta}_1^*})^\T}{\partial \bm{c}_0}\Bigg|_{\bm{\theta} = \widehat{\bm{\theta}}} \text{cov}(\widehat{\bm{c}}_0, \widehat{\bm{C}}_1^\T) \frac{\partial \bm{\beta}_1}{\partial \bm{C}_1^\T}\Bigg|_{\bm{\theta} = \widehat{\bm{\theta}}}\\
        &- 
		\frac{\partial (\bm{C}\bm{B}{\bm{\beta}_1^*})^\T}{\partial \bm{\Sigma}_e}\Bigg|_{\bm{\theta} = \widehat{\bm{\theta}}} \text{var}(\widehat{\bm{\Sigma}}_e) \frac{\partial \bm{\beta}_1}{\partial \bm{\Sigma}_e}\Bigg|_{\bm{\theta} = \widehat{\bm{\theta}}} -
		\frac{\partial (\bm{C}\bm{B}{\bm{\beta}_1^*})^\T}{\partial \bm{\Sigma}_{z}}\Bigg|_{\bm{\theta} = \widehat{\bm{\theta}}} \text{var}(\widehat{\bm{\Sigma}}_{z}) \frac{\partial \bm{\beta}_1}{\partial \bm{\Sigma}_{z}}\Bigg|_{\bm{\theta} = \widehat{\bm{\theta}}}.
	\end{aligned}
\end{equation*}

For the covariance between $\widehat{\beta}_0$ and $\widehat{\bm{\beta}}_2$:
\begin{equation*}
	\begin{aligned}
		\text{cov}(\widehat{\beta}_0, \widehat{\bm{\beta}}_2) &= \text{cov}(\widehat{\beta}_0^*, \widehat{\bm{\beta}}_2^*) - \widehat{\bm{C}}\widehat{\bm{B}}\text{cov}(\widehat{\bm{\beta}}_2^*, \widehat{\bm{\beta}}_1^*)  - \widehat{\bm{D}}\widehat{\bm{B}}\text{cov}(\widehat{\beta}_0^*, \widehat{\bm{\beta}}_1^*)
        +  \frac{\partial (\bm{C}\bm{B}{\bm{\beta}_1^*})^\T}{\partial \bm{\beta}_1^*}\Bigg|_{\bm{\theta} = \widehat{\bm{\theta}}} \text{var}(\widehat{\bm{\beta}}_1^*) \frac{\partial (\bm{D}\bm{B}{\bm{\beta}_1^*})}{\partial \bm{\beta}_1^*}\Bigg|_{\bm{\theta} = \widehat{\bm{\theta}}}\\
        &+ 
		\frac{\partial (\bm{C}\bm{B}{\bm{\beta}_1^*})^\T}{\partial \bm{c}_0}\Bigg|_{\bm{\theta} = \widehat{\bm{\theta}}} \text{cov}(\widehat{\bm{c}}_0, \widehat{\bm{C}}_2^\T) \frac{\partial (\bm{D}\bm{B}{\bm{\beta}_1^*})}{\partial \bm{C}_2^\T}\Bigg|_{\bm{\theta} = \widehat{\bm{\theta}}} + \frac{\partial (\bm{C}\bm{B}{\bm{\beta}_1^*})^\T}{\partial \bm{b}_{0}}\Bigg|_{\bm{\theta} = \widehat{\bm{\theta}}}\text{cov}(\widehat{\bm{b}}_0, \widehat{\bm{B}}_2^\T)\frac{\partial (\bm{D}\bm{B}{\bm{\beta}_1^*})}{\partial \bm{B}_2^\T}\Bigg|_{\bm{\theta} = \widehat{\bm{\theta}}}\\
        &+ 
		\frac{\partial (\bm{C}\bm{B}{\bm{\beta}_1^*})^\T}{\partial \bm{\Sigma}_e}\Bigg|_{\bm{\theta} = \widehat{\bm{\theta}}} \text{var}(\widehat{\bm{\Sigma}}_e) \frac{\partial (\bm{D}\bm{B}{\bm{\beta}_1^*})}{\partial \bm{\Sigma}_e}\Bigg|_{\bm{\theta} = \widehat{\bm{\theta}}} +
		\frac{\partial (\bm{C}\bm{B}{\bm{\beta}_1^*})^\T}{\partial \bm{\Sigma}_{z}}\Bigg|_{\bm{\theta} = \widehat{\bm{\theta}}} \text{var}(\widehat{\bm{\Sigma}}_{z}) \frac{\partial (\bm{D}\bm{B}{\bm{\beta}_1^*})}{\partial \bm{\Sigma}_{z}}\Bigg|_{\bm{\theta} = \widehat{\bm{\theta}}}.
	\end{aligned}
\end{equation*}

For the covariance between $\widehat{\bm{\beta}}_1$ and $\widehat{\bm{\beta}}_2$:
\begin{equation*}
	\begin{aligned}
		\text{cov}(\widehat{\bm{\beta}}_1, \widehat{\bm{\beta}}_2) &=  \widehat{\bm{A}}\widehat{\bm{B}}\text{cov}(\widehat{\bm{\beta}}_2^*, \widehat{\bm{\beta}}_1^*) 
        -  \frac{\partial \bm{\beta}_1^\T}{\partial \bm{\beta}_1^*}\Bigg|_{\bm{\theta} = \widehat{\bm{\theta}}} \text{var}(\widehat{\bm{\beta}}_1^*) \frac{\partial (\bm{D}\bm{B}{\bm{\beta}_1^*})}{\partial \bm{\beta}_1^*}\Bigg|_{\bm{\theta} = \widehat{\bm{\theta}}}
        - 
		\frac{\partial \bm{\beta}_1^\T}{\partial \bm{C}_1^\T}\Bigg|_{\bm{\theta} = \widehat{\bm{\theta}}} \text{cov}(\widehat{\bm{C}}_1^\T, \widehat{\bm{C}}_2^\T) \frac{\partial (\bm{D}\bm{B}{\bm{\beta}_1^*})}{\partial \bm{C}_2^\T}\Bigg|_{\bm{\theta} = \widehat{\bm{\theta}}}\\
        &- 
		\frac{\partial \bm{\beta}_1^\T}{\partial \bm{\Sigma}_e}\Bigg|_{\bm{\theta} = \widehat{\bm{\theta}}} \text{var}(\widehat{\bm{\Sigma}}_e) \frac{\partial (\bm{D}\bm{B}{\bm{\beta}_1^*})}{\partial \bm{\Sigma}_e}\Bigg|_{\bm{\theta} = \widehat{\bm{\theta}}} -
		\frac{\partial \bm{\beta}_1^\T}{\partial \bm{\Sigma}_{z}}\Bigg|_{\bm{\theta} = \widehat{\bm{\theta}}} \text{var}(\widehat{\bm{\Sigma}}_{z}) \frac{\partial (\bm{D}\bm{B}{\bm{\beta}_1^*})}{\partial \bm{\Sigma}_{z}}\Bigg|_{\bm{\theta} = \widehat{\bm{\theta}}}.
	\end{aligned}
\end{equation*}

In addition, $\text{var}(\widehat{\beta}_0^*)$, $\text{var}(\widehat{\bm{\beta}}_1^*)$, $\text{var}(\widehat{\bm{\beta}}_2^*)$, $\text{cov}(\widehat{\beta}_0^*, \widehat{\bm{\beta}}_1^*)$, $\text{cov}(\widehat{\beta}_0^*, \widehat{\bm{\beta}}_2^*)$ and $\text{cov}(\widehat{\bm{\beta}}_2^*, \widehat{\bm{\beta}}_1^*)$ are obtained from fitting \eqref{C1} to the main study data,  $\text{var}(\widehat{\bm{c}}_0)$, $\text{var}(\widehat{\bm{C}}_1^\T)$, $\text{var}(\widehat{\bm{C}}_2^\T)$, $\text{cov}(\widehat{\bm{c}}_0, \widehat{\bm{C}}_1^\T)$, $\text{cov}(\widehat{\bm{c}}_0, \widehat{\bm{C}}_2^\T)$, $\text{cov}(\widehat{\bm{C}}_1^\T, \widehat{\bm{C}}_2^\T)$ and $\text{var}(\widehat{\bm{\Sigma}}_e)$ are obtained from fitting \eqref{C2} to the validation study data, $\text{var}(\widehat{\bm{b}}_{0})$, $\text{var}(\widehat{\bm{B}}_{2}^\T)$, $\text{cov}(\widehat{\bm{b}}_0, \widehat{\bm{B}}_2^\T)$ and $\text{var}(\widehat{\bm{\Sigma}}_{z})$ can be obtained from fitting \eqref{C3} to the main study data based on \cite{Fisher1930}. 

Finally, $\surd{n_M}(\widehat{\bm{\beta}}-\bm{\beta})$ is asymptotically mean-zero multivariate normal with covariance matrix
\begin{equation*}
		n_M\begin{pmatrix}
			\text{var}(\widehat{{\beta}}_0) & \text{cov}(\widehat{\beta}_0, \widehat{\bm{\beta}}_1) & \text{cov}(\widehat{\beta}_0, \widehat{\bm{\beta}}_2) \\
			\text{cov}(\widehat{\beta}_0, \widehat{\bm{\beta}}_1) & \text{var}(\widehat{\bm{\beta}}_1) & \text{cov}(\widehat{\bm{\beta}}_1, \widehat{\bm{\beta}}_2) \\
			\text{cov}(\widehat{\beta}_0, \widehat{\bm{\beta}}_2) & \text{cov}(\widehat{\bm{\beta}}_1, \widehat{\bm{\beta}}_2)& \text{var}(\widehat{\bm{\beta}}_2)
		\end{pmatrix},
\end{equation*} 
when both $n_M$ and $n_V$ approach infinity and $n_M/n_V \rightarrow \lambda$, where $0<\lambda<\infty$.  

\section*{Appendix 2: More results for simulation studies}

As with the models in the main text, there are
\begin{equation}
	\bm{X}_i = \bm{a}_0 + \bm{A}_2^\T \bm{W}_i + \bm{\epsilon}.\label{D1}
\end{equation}

In practice, the transformed form of exposures is sometimes used to preserve the normality of $\bm{\epsilon}_e$ in \eqref{C2}; for example, \cite{Subar2003} applied log transformations to the exposures to achieve this. However, we cannot guarantee that $\bm{\epsilon}$ in \eqref{D1} also follows the normal distribution simultaneously. Hence, we drop the normality assumption of this error term when generating true exposures (i.e. $\bm{X}\mid{W}$ is not normally distributed), to examine the robustness of our proposed method. 

\begin{figure}[!htbp]
    \includegraphics[scale=0.6]{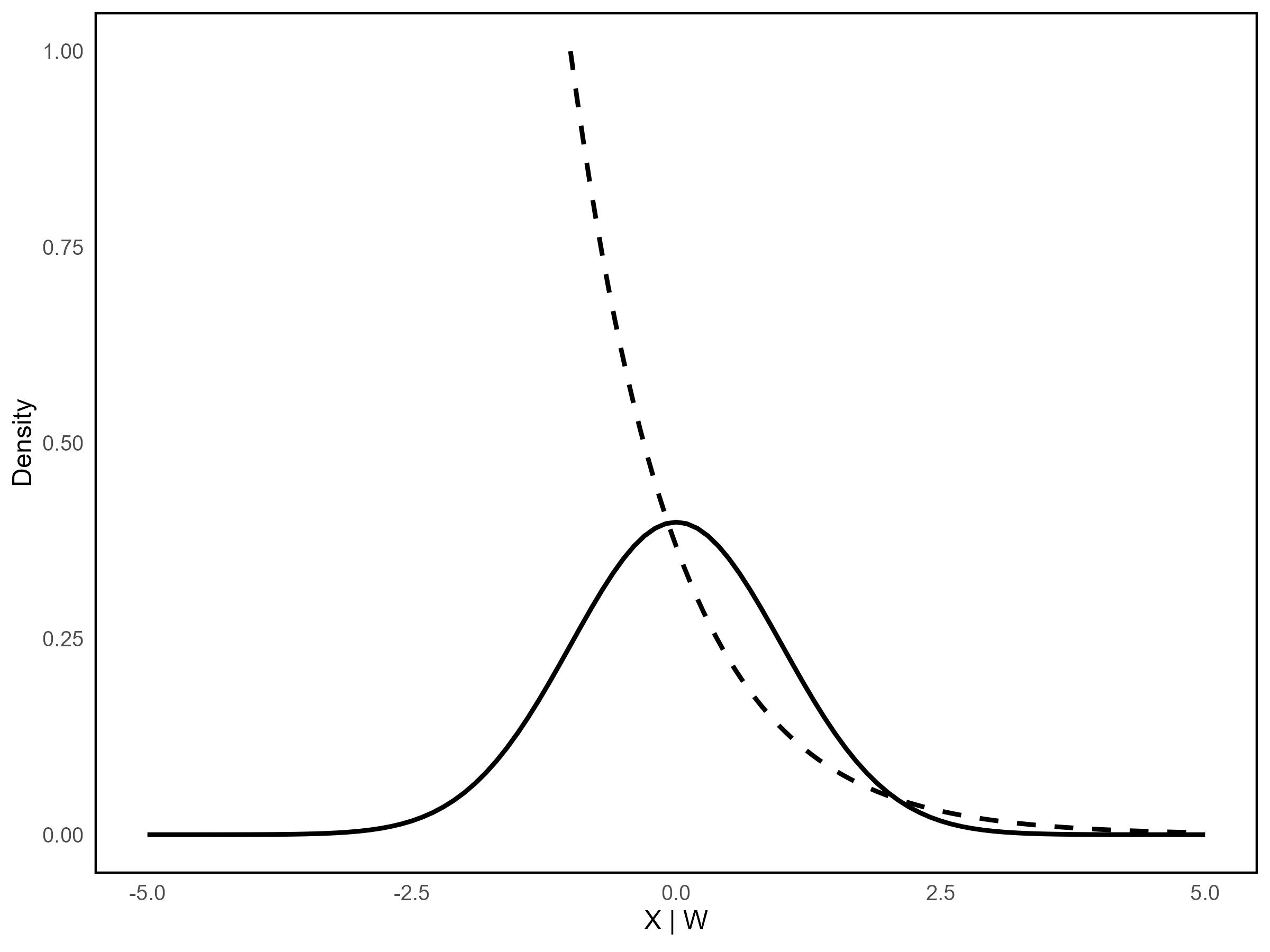}
    \caption{A graph showing the normal distribution (solid), and the gamma distribution (dashes).}
    \label{fig1}
\end{figure}

Settings for the sample size and single error-free confounder were unchanged compared with the main text. Specifically, $n_M=10,000$ for the main study, $n_V=500$ for the external validation study, and $W$ followed a normal distribution $\mathcal{N}(1,1)$. For the single-exposure case, we substituted the distribution of $X\mid{W}$ with a gamma distribution, which considerably differs from the normal distribution while maintaining the mean and covariance of ${X}\mid{W}$ same as the simulation studies. The gamma distribution had a rate parameter of 1 and was shifted to have a mean of 0. The plots of normal and gamma distributions are shown in Fig.~\ref{fig1}. 

For the multiple-exposure case, $\bm{X}\mid{W}$ were generated to have gamma marginals. To preserve the covariance structure between variables, we converted normally distributed $\bm{X}\mid{W}$ in the main text to uniform marginals, then applied the inverse gamma to each uniform marginal to get the final distribution keeping the mean and covariance in the main text. For each marginal distribution of $\bm{X}\mid{W}$, its rate is 1, and the difference from the normal distribution is similar to that in Fig.~\ref{fig1}.

Then as same as the main text, model \eqref{C2} and model \eqref{C1} were used to generate surrogate exposures $\bm{Z}$ and the outcome $\bm{Y}$. Three scenarios of true exposures $\bm{X}$ were conducted, where the first satisfies the double transportability assumption, while the other two do not. In each scenario, using the rule of thumb suggested by \cite{Kuha1994}, we investigated two sets of parameters to represent small and large measurement errors, respectively. More detailed, $\beta_1 = 1$ in single-exposure case and $\bm{\beta}_1 = (1.2, 1.1, 0.9, 0.8)^\T$ in multiple-exposure case were used, and $\bm{\Sigma}_e$ was varied to mimic small and large measurement error level, respectively. 

The simulations were repeated 10,000 times for each scenario, yielding the following results. Similar findings were observed in the main text, which implies our proposed method is robust to different error distributions.

\begin{table}[!htbp]
	\centering
	\caption{Simulation results for the single-exposure case with $\beta_1 = 1$ and gamma distributions. ME is the measurement error level. $\widehat{\beta}_1$ is the estimate of $\beta_1$. Bias(\%) is the relative bias, i.e. Bias(\%)=$(\widehat{\beta}_1 - \beta_1)/\beta_1\times 100$. SE is the mean of the estimated standard error of $\widehat{\beta}_1$. SD is the empirical standard deviation of the $\widehat{\beta}_1$ values. CP is the empirical coverage probability of the asymptotic 95\% confidence interval.}
	\begin{tabular}{cccccccc}
\toprule
		& ME  & Method & $\widehat{\beta}_1$ & Bias(\%)   & SE    & SD    & CP \\ \midrule
		\multirow{6}[0]{*}{\rotatebox{90}{Scenario 1}} & \multirow{3}[0]{*}{\rotatebox{90}{Small}}   & Our proposed method & 1.00  & -0.13\% & 0.047 & 0.047 & 94.80\% \\
          &       & Regression calibration & 1.01  & 0.55\% & 0.049 & 0.051 & 94.23\% \\
          &       & Naive estimator & 0.67  & -32.89\% & 0.009 & 0.011 & 0.00\% \\
 \cmidrule{2-8}
		& \multirow{3}[0]{*}{\rotatebox{90}{Large}}    & Our proposed method & 1.01  & 0.90\% & 0.151 & 0.154 & 93.41\% \\
          &       & Regression calibration & 1.01  & 1.21\% & 0.098 & 0.101 & 94.29\% \\
          &       & Naive estimator & 0.34  & -66.21\% & 0.007 & 0.009 & 0.00\% \\ \midrule
		\multirow{6}[0]{*}{\rotatebox{90}{Scenario 2}} & \multirow{3}[0]{*}{\rotatebox{90}{Small}}   & Our proposed method & 1.00  & -0.12\% & 0.050 & 0.050 & 94.53\% \\
          &       & Regression calibration & 1.09  & 8.94\% & 0.063 & 0.067 & 72.96\% \\
          &       & Naive estimator & 0.67  & -32.90\% & 0.009 & 0.011 & 0.00\% \\ \cmidrule{2-8}
		& \multirow{3}[0]{*}{\rotatebox{90}{Large}}     & Our proposed method & 1.01  & 0.65\% & 0.153 & 0.153 & 93.85\% \\
          &       & Regression calibration & 1.18  & 18.47\% & 0.132 & 0.140 & 76.01\% \\
          &       & Naive estimator & 0.34  & -66.22\% & 0.007 & 0.009 & 0.00\% \\ \midrule
		\multirow{6}[0]{*}{\rotatebox{90}{Scenario 3}} & \multirow{3}[0]{*}{\rotatebox{90}{Small}}   & Our proposed method & 1.00  & -0.04\% & 0.045 & 0.045 & 94.60\% \\
          &       & Regression calibration & 0.94  & -6.26\% & 0.039 & 0.040 & 60.58\% \\
          &       & Naive estimator & 0.67  & -32.90\% & 0.009 & 0.011 & 0.00\% \\ \cmidrule{2-8}
		& \multirow{3}[0]{*}{\rotatebox{90}{Large}}     & Our proposed method & 1.01  & 0.89\% & 0.147 & 0.148 & 93.85\% \\
          &       & Regression calibration & 0.88  & -12.42\% & 0.074 & 0.077 & 56.21\% \\
          &       & Naive estimator & 0.34  & -66.20\% & 0.007 & 0.009 & 0.00\% \\
  \bottomrule
	\end{tabular}%
	\label{tc}%
        \tabnote{\\Scenario 1 satisfies the double transportability assumption, i.e. $\bm{\mu}_M=\bm{\mu}_V, \bm{\Sigma}_M=\bm{\Sigma}_V$. While Scenarios 2 and 3 only meet the single transportability assumption, i.e. $\bm{\mu}_V = 0.8\bm{\mu}_M, \bm{\Sigma}_V=0.8\bm{\Sigma}_M$ for Scenario 2 and $\bm{\mu}_V = 1.25\bm{\mu}_M, \bm{\Sigma}_V = 1.25\bm{\Sigma}_M$ for Scenario 3. }
\end{table}%

\begin{table}[!htbp]
  \centering
  \caption{Simulation results for the multiple-exposure case with small measurement errors. $\beta$ is the true value of the corresponding component of $\bm{\beta}_1$. 
    $\widehat{\beta}$ is the estimate of $\beta$. Bias(\%) is the relative bias, i.e. Bias(\%)=$(\widehat{\beta} - \beta)/\beta\times 100$. SE is the mean of the estimated standard error of $\widehat{\beta}$. SD is the empirical standard deviation of the $\widehat{\beta}$ values. CP is the empirical coverage probability of the asymptotic 95\% confidence interval.}
    \begin{tabular}{ccccccccc}
    \toprule
          &  & $\beta$  & Method & $\widehat{\beta}$ & Bias(\%)   & SE    & SD    & CP \\
    \midrule
        \multirow{12}[0]{*}{\rotatebox{90}{Scenario 1}} & $\beta_{1,1}$ & 1.2   & Our proposed method & 1.20  & -0.17\% & 0.077 & 0.077 & 95.04\% \\
          &       &       & Regression calibration & 1.20  & 0.34\% & 0.066 & 0.076 & 91.06\% \\
          &       &       & Naive estimator & 0.73  & -39.45\% & 0.013 & 0.014 & 0.00\% \\\cmidrule{2-9}
          & $\beta_{1,2}$ & 1.1   & Our proposed method & 1.10  & -0.27\% & 0.076 & 0.076 & 94.79\% \\
          &       &       & Regression calibration & 1.10  & 0.25\% & 0.066 & 0.075 & 91.91\% \\
          &       &       & Naive estimator & 0.66  & -40.25\% & 0.013 & 0.014 & 0.00\% \\\cmidrule{2-9}
          & $\beta_{1,3}$ & 0.9   & Our proposed method & 0.90  & -0.05\% & 0.075 & 0.075 & 95.39\% \\
          &       &       & Regression calibration & 0.90  & 0.43\% & 0.066 & 0.071 & 93.08\% \\
          &       &       & Naive estimator & 0.52  & -42.35\% & 0.013 & 0.013 & 0.00\% \\\cmidrule{2-9}
          & $\beta_{1,4}$ & 0.8   & Our proposed method & 0.80  & -0.13\% & 0.075 & 0.075 & 95.65\% \\
          &       &       & Regression calibration & 0.80  & 0.27\% & 0.066 & 0.070 & 93.26\% \\
          &       &       & Naive estimator & 0.45  & -43.78\% & 0.013 & 0.013 & 0.00\% \\
    \midrule
        \multirow{12}[0]{*}{\rotatebox{90}{Scenario 2}} & $\beta_{1,1}$ & 1.2   & Our proposed method & 1.20  & -0.16\% & 0.084 & 0.085 & 94.55\% \\
          &       &       & Regression calibration & 1.31  & 9.26\% & 0.083 & 0.098 & 72.61\% \\
          &       &       & Naive estimator & 0.73  & -39.47\% & 0.013 & 0.014 & 0.00\% \\\cmidrule{2-9}
          & $\beta_{1,2}$ & 1.1   & Our proposed method & 1.10  & -0.09\% & 0.084 & 0.084 & 95.07\% \\
          &       &       & Regression calibration & 1.20  & 9.00\% & 0.083 & 0.095 & 77.71\% \\
          &       &       & Naive estimator & 0.66  & -40.22\% & 0.013 & 0.013 & 0.00\% \\\cmidrule{2-9}
          & $\beta_{1,3}$ & 0.9   & Our proposed method & 0.90  & -0.05\% & 0.083 & 0.082 & 95.39\% \\
          &       &       & Regression calibration & 0.97  & 7.93\% & 0.083 & 0.090 & 85.42\% \\
          &       &       & Naive estimator & 0.52  & -42.33\% & 0.013 & 0.013 & 0.00\% \\\cmidrule{2-9}
          & $\beta_{1,4}$ & 0.8   & Our proposed method & 0.80  & -0.15\% & 0.082 & 0.083 & 95.15\% \\
          &       &       & Regression calibration & 0.86  & 6.90\% & 0.083 & 0.089 & 89.03\% \\
          &       &       & Naive estimator & 0.45  & -43.76\% & 0.013 & 0.013 & 0.00\% \\
    \midrule
        \multirow{12}[0]{*}{\rotatebox{90}{Scenario 3}} & $\beta_{1,1}$ & 1.2   & Our proposed method & 1.20  & -0.20\% & 0.071 & 0.072 & 94.70\% \\
          &       &       & Regression calibration & 1.12  & -6.49\% & 0.052 & 0.061 & 63.09\% \\
          &       &       & Naive estimator & 0.73  & -39.45\% & 0.013 & 0.014 & 0.00\% \\\cmidrule{2-9}
          & $\beta_{1,2}$ & 1.1   & Our proposed method & 1.10  & -0.20\% & 0.070 & 0.070 & 95.28\% \\
          &       &       & Regression calibration & 1.03  & -6.18\% & 0.052 & 0.059 & 69.58\% \\
          &       &       & Naive estimator & 0.66  & -40.26\% & 0.013 & 0.013 & 0.00\% \\\cmidrule{2-9}
          & $\beta_{1,3}$ & 0.9   & Our proposed method & 0.90  & -0.09\% & 0.069 & 0.068 & 95.22\% \\
          &       &       & Regression calibration & 0.85  & -5.46\% & 0.052 & 0.058 & 79.81\% \\
          &       &       & Naive estimator & 0.52  & -42.34\% & 0.013 & 0.013 & 0.00\% \\\cmidrule{2-9}
          & $\beta_{1,4}$ & 0.8   & Our proposed method & 0.80  & -0.23\% & 0.069 & 0.069 & 95.14\% \\
          &       &       & Regression calibration & 0.76  & -4.94\% & 0.052 & 0.056 & 85.07\% \\
          &       &       & Naive estimator & 0.45  & -43.79\% & 0.013 & 0.013 & 0.00\% \\
    \bottomrule
    \end{tabular}%
    \tabnote{\\Scenario 1 satisfies the double transportability assumption, i.e. $\bm{\mu}_M=\bm{\mu}_V, \bm{\Sigma}_M=\bm{\Sigma}_V$. While Scenarios 2 and 3 only meet the single transportability assumption, i.e. $\bm{\mu}_V = 0.8\bm{\mu}_M, \bm{\Sigma}_V=0.8\bm{\Sigma}_M$ for Scenario 2 and $\bm{\mu}_V = 1.25\bm{\mu}_M, \bm{\Sigma}_V = 1.25\bm{\Sigma}_M$ for Scenario 3. }
  \label{multi_small_gamma}
\end{table}%

\begin{table}[!htbp]
  \centering
  \caption{Simulation results for the multiple-exposure case with large measurement errors. $\beta$ is the true value of the corresponding component of $\bm{\beta}_1$. $\widehat{\beta}$ is the estimate of $\beta$. Bias(\%) is the relative bias, i.e. Bias(\%)=$(\widehat{\beta} - \beta)/\beta\times 100$. SE is the mean of the estimated standard error of $\widehat{\beta}$. SD is the empirical standard deviation of the $\widehat{\beta}$ values. CP is the empirical coverage probability of the asymptotic 95\% confidence interval.}
    \begin{tabular}{ccccccccc}
    \toprule
		&  & $\beta$  & Method & $\widehat{\beta}$ & Bias(\%)   & SE    & SD    & CP \\
    \midrule
    \multirow{12}[0]{*}{\rotatebox{90}{Scenario 1}} & $\beta_{1,1}$ & 1.2   & Our proposed method & 1.21  & 0.64\% & 0.188 & 0.185 & 95.75\% \\
          &       &       & Regression calibration & 1.21  & 0.77\% & 0.123 & 0.147 & 90.33\% \\
          &       &       & Naive estimator & 0.48  & -60.10\% & 0.012 & 0.014 & 0.00\% \\\cmidrule{2-9}
          & $\beta_{1,2}$ & 1.1   & Our proposed method & 1.11  & 0.62\% & 0.186 & 0.181 & 96.33\% \\
          &       &       & Regression calibration & 1.11  & 0.74\% & 0.123 & 0.140 & 91.99\% \\
          &       &       & Naive estimator & 0.44  & -60.05\% & 0.012 & 0.013 & 0.00\% \\\cmidrule{2-9}
          & $\beta_{1,3}$ & 0.9   & Our proposed method & 0.90  & 0.55\% & 0.181 & 0.177 & 96.77\% \\
          &       &       & Regression calibration & 0.91  & 0.88\% & 0.123 & 0.135 & 92.60\% \\
          &       &       & Naive estimator & 0.36  & -59.96\% & 0.012 & 0.013 & 0.00\% \\\cmidrule{2-9}
          & $\beta_{1,4}$ & 0.8   & Our proposed method & 0.80  & 0.31\% & 0.180 & 0.177 & 96.65\% \\
          &       &       & Regression calibration & 0.80  & 0.60\% & 0.123 & 0.134 & 92.64\% \\
          &       &       & Naive estimator & 0.32  & -59.87\% & 0.012 & 0.013 & 0.00\% \\
    \midrule
    \multirow{12}[0]{*}{\rotatebox{90}{Scenario 2}} & $\beta_{1,1}$ & 1.2   & Our proposed method & 1.21  & 0.77\% & 0.198 & 0.195 & 95.95\% \\
          &       &       & Regression calibration & 1.42  & 18.39\% & 0.169 & 0.198 & 74.80\% \\
          &       &       & Naive estimator & 0.48  & -60.10\% & 0.012 & 0.014 & 0.00\% \\\cmidrule{2-9}
          & $\beta_{1,2}$ & 1.1   & Our proposed method & 1.11  & 0.79\% & 0.195 & 0.194 & 95.84\% \\
          &       &       & Regression calibration & 1.30  & 17.88\% & 0.169 & 0.194 & 79.34\% \\
          &       &       & Naive estimator & 0.44  & -60.05\% & 0.012 & 0.013 & 0.00\% \\\cmidrule{2-9}
          & $\beta_{1,3}$ & 0.9   & Our proposed method & 0.91  & 0.57\% & 0.191 & 0.186 & 96.53\% \\
          &       &       & Regression calibration & 1.05  & 16.83\% & 0.169 & 0.187 & 84.93\% \\
          &       &       & Naive estimator & 0.36  & -59.93\% & 0.012 & 0.013 & 0.00\% \\\cmidrule{2-9}
          & $\beta_{1,4}$ & 0.8   & Our proposed method & 0.80  & 0.20\% & 0.190 & 0.188 & 96.52\% \\
          &       &       & Regression calibration & 0.92  & 15.28\% & 0.169 & 0.182 & 88.55\% \\
          &       &       & Naive estimator & 0.32  & -59.86\% & 0.012 & 0.013 & 0.00\% \\
    \midrule
    \multirow{12}[0]{*}{\rotatebox{90}{Scenario 3}} & $\beta_{1,1}$ & 1.2   & Our proposed method & 1.21  & 0.57\% & 0.180 & 0.177 & 96.04\% \\
          &       &       & Regression calibration & 1.05  & -12.67\% & 0.090 & 0.109 & 55.56\% \\
          &       &       & Naive estimator & 0.48  & -60.10\% & 0.012 & 0.014 & 0.00\% \\\cmidrule{2-9}
          & $\beta_{1,2}$ & 1.1   & Our proposed method & 1.11  & 0.66\% & 0.177 & 0.173 & 96.29\% \\
          &       &       & Regression calibration & 0.97  & -12.16\% & 0.090 & 0.107 & 61.75\% \\
          &       &       & Naive estimator & 0.44  & -60.06\% & 0.012 & 0.013 & 0.00\% \\\cmidrule{2-9}
          & $\beta_{1,3}$ & 0.9   & Our proposed method & 0.90  & 0.52\% & 0.172 & 0.167 & 96.55\% \\
          &       &       & Regression calibration & 0.80  & -11.44\% & 0.090 & 0.104 & 72.62\% \\
          &       &       & Naive estimator & 0.36  & -59.96\% & 0.012 & 0.013 & 0.00\% \\\cmidrule{2-9}
          & $\beta_{1,4}$ & 0.8   & Our proposed method & 0.80  & -0.02\% & 0.171 & 0.166 & 96.83\% \\
          &       &       & Regression calibration & 0.71  & -10.78\% & 0.090 & 0.099 & 78.95\% \\
          &       &       & Naive estimator & 0.32  & -59.89\% & 0.012 & 0.013 & 0.00\% \\
    \bottomrule
    \end{tabular}%
  \label{multi_large_gamma}%
  \tabnote{\\Scenario 1 satisfies the double transportability assumption, i.e. $\bm{\mu}_M=\bm{\mu}_V, \bm{\Sigma}_M=\bm{\Sigma}_V$. While Scenarios 2 and 3 only meet the single transportability assumption, i.e. $\bm{\mu}_V = 0.8\bm{\mu}_M, \bm{\Sigma}_V=0.8\bm{\Sigma}_M$ for Scenario 2 and $\bm{\mu}_V = 1.25\bm{\mu}_M, \bm{\Sigma}_V = 1.25\bm{\Sigma}_M$ for Scenario 3. }
\end{table}%

\end{document}